\documentclass[twocolumn]{aastex63}
\usepackage{times,epsfig,amsmath}
\usepackage[T1]{fontenc}

\hypersetup{pdfauthor={Ewan O'Sullivan},
            pdftitle={The cluster-central compact steep-spectrum radio galaxy 1321+045},
            pdfkeywords={galaxies: clusters: individual (MaxBCG J201.08197+04.31863), galaxies: individual (SDSS J132419.67+041907.1), galaxies: active, X-rays: galaxies: clusters, galaxies: clusters: intracluster medium},
            bookmarksnumbered=true}
\pdfoutput=1

\newcommand{\arcm}{\hbox{$^\prime$}}

\newcommand{\degree}{\hbox{$^\circ$}}

\newcommand{\chandra}{\emph{Chandra}}

\newcommand{\arcs}{\mbox{\arcm\arcm}}

\newcommand{\Zsol}{\ensuremath{\mathrm{~Z_{\odot}}}}

\newcommand{\Msol}{\ensuremath{\mathrm{~M_{\odot}}}}
\newcommand{\Msolpyr}{\ensuremath{\mathrm{~M_{\odot}~yr^{-1}}}}

\newcommand{\s}{\ensuremath{\mbox{~s}}}
\newcommand{\ps}{\ensuremath{\s^{-1}}}
\newcommand{\cm}{\ensuremath{\mbox{~cm}}}
\newcommand{\pcmsq}{\ensuremath{\cm^{-2}}}

\newcommand{\kev}{\ensuremath{\mbox{~keV}}}
\newcommand{\kevcmsq}{\ensuremath{\kev\cm^{2}}}
\newcommand{\km}{\ensuremath{\mbox{~km}}}
\newcommand{\Mpc}{\ensuremath{\mbox{~Mpc}}}
\newcommand{\pMpc}{\ensuremath{\Mpc^{-1}}}
\newcommand{\kmpspMpc}{\ensuremath{\km \ps \pMpc\,}}
\newcommand{\erg}{\ensuremath{\mbox{~erg}}}
\newcommand{\ergps}{\ensuremath{\erg \ps}}

\newcommand{\kmps}{\ensuremath{\km \ps}}

\newcommand{\Dtf}{\ensuremath{D_{\mathrm{25}}}}

\newcommand{\gtsim}{\,\rlap{\raise 0.5ex\hbox{$>$}}{\lower 1.0ex\hbox{$\sim$}}\,}

\newcommand{\MaxBCG}{MaxBCG~J201.08197+04.31863}

\received{2021 Feb 15}
\revised{2021 Mar 23}
\accepted{2021 Apr 9}
\submitjournal{ApJ}

\shorttitle{Cluster-central CSS 1321+045}
\shortauthors{O'Sullivan et al.}

\begin{document}

\title{
The cluster-central compact steep-spectrum radio galaxy 1321+045
}

\correspondingauthor{Ewan O'Sullivan}
\email{eosullivan@cfa.harvard.edu}

\author[0000-0002-5671-6900]{Ewan O'Sullivan}
\affiliation{Center for Astrophysics $|$ Harvard \& Smithsonian, 60 Garden
  Street, Cambridge, MA 02138, USA}

\author[0000-0002-6741-9856]{Magdalena Kunert-Bajraszewska}
\affiliation{Institute of Astronomy, Faculty of Physics, Astronomy and Informatics, NCU, Grudzi\k{a}dzka 5, 87-100 Toru\'{n}, Poland}

\author[0000-0002-0905-7375]{Aneta Siemiginowska}
\affiliation{Center for Astrophysics $|$ Harvard \& Smithsonian, 60 Garden
  Street, Cambridge, MA 02138, USA}

\author[0000-0003-4428-7835]{D. J. Burke}
\affiliation{Center for Astrophysics $|$ Harvard \& Smithsonian, 60 Garden
  Street, Cambridge, MA 02138, USA}

\author[0000-0003-2658-7893]{Fran\c{c}oise Combes}
\affiliation{LERMA, Observatoire de Paris, CNRS, PSL Univ., Sorbonne Univ., 75014 Paris, France}
\affiliation{Coll\`{e}ge de France, 11 place Marcelin Berthelot, 75005 Paris, France}

\author{Philippe Salom\'{e}}
\affiliation{LERMA, Observatoire de Paris, CNRS, PSL Univ., Sorbonne Univ., 75014 Paris, France}

\author[0000-0002-1634-9886]{Simona Giacintucci}
\affiliation{Naval Research Laboratory, 4555 Overlook Avenue SW, Code 7213, Washington, DC 20375, USA}

\begin{abstract}
The radio galaxy 1321+045 is a rare example of a young, compact steep spectrum source located in the center of a $z$=0.263 galaxy cluster. Using a combination of \chandra, VLBA, VLA, MERLIN and IRAM~30m observations, we investigate the  conditions which have triggered this outburst. We find that the previously identified 5~kpc scale radio lobes are probably no longer powered by the AGN, which seems to have launched a new $\sim$20~pc jet on a different axis, likely within the last few hundred years. We estimate the enthalpy of the lobes to be 8.48$^{+6.04}_{-3.56}\times10^{57}$~erg, only sufficient to balance cooling in the surrounding 16~kpc for $\sim$9~Myr. The cluster ICM properties are similar to those of rapidly cooling nearby clusters, with a low central entropy (8.6$^{+2.2}_{-1.4}$\kevcmsq\ within 8~kpc), short central cooling time (390$^{+170}_{-150}$~Myr), and t$_{cool}$/t$_{ff}$ and t$_{cool}$/t$_{eddy}$ ratios indicative of thermal instability out to $\sim$45~kpc. Despite previous detection of H$\alpha$ emission from the BCG, our IRAM~30m observations do not detect CO emission in either the (1-0) or (3-2) transitions. We place 3$\sigma$ limits on the molecular gas mass of M$_{\rm mol}\leq$7.7$\times$10$^9$\Msol\ and $\leq$5.6$\times$10$^9$\Msol\ from the two lines respectively. We find indications of a recent minor cluster merger which has left a $\sim$200~kpc tail of stripped gas in the ICM, and probably induced sloshing motions.
\end{abstract}


\section{Introduction}
\label{sec:intro}

The dominant galaxies of cool-core galaxy groups and clusters often host powerful radio galaxies, most commonly Fanaroff-Riley type I systems \citep[FR~Is,][]{FanaroffRiley74}. Strong evidence \citep{Sunetal09} links radio activity in these central galaxies with cooling in the intra-cluster medium (ICM) and with the presence of molecular and ionized gas, thought to have cooled from the ICM in and around those galaxies \citep[e.g.,][]{Lakhchauraetal18,Babyketal19}. X-ray observations reveal many ICM structures associated with the radio jets and lobes, including cavities, shocks and ripples \citep[e.g.,][]{Fabian12,McNamaraNulsen12,Gittietal12} and it has been shown that the energies required to create these structures are in general closely correlated with energy losses from the ICM through radiative cooling \citep[e.g.,][]{Birzanetal04,Raffertyetal06,Panagouliaetal14b}. It is now widely accepted that, at least at low redshift, cool core systems operate in a feedback loop, in which the active galactic nucleus (AGN) of the dominant galaxy regulates the cooling of the entire halo, through cycles of activity lasting tens to hundreds of Myr.

While the lifespan of FR~I (and FR~II) radio galaxies may be a few tens of Myr, compact steep spectrum (CSS) and gigahertz peak spectrum (GPS) radio sources are thought to be much younger, $<$10$^5$~yr old \citep{Readheadetal96a, Fantietal95}. They may be at the start of their development into much larger lobed or plumed radio galaxies \citep{Readheadetal96b}. As such they offer an opportunity to study the earliest phases of the AGN heating cycle, in terms of the evolution of the radio source, its impact on its immediate surroundings, and perhaps most interestingly in terms of the ICM conditions which trigger activity. This would complement studies of clusters and groups with older central FR~Is \citep[e.g.,][]{Hoganetal17,Lakhchauraetal18,Olivaresetal19} which provide information on ICM cooling and AGN fuelling but only after the feedback has begun to affect the gas.

However, as is to be expected for such a short-lived phase of activity, CSS and GPS sources in cluster-dominant galaxies are rare, with only a handful of known examples. 3C~186 is probably the best studied to date, a relaxed cool core cluster at redshift $z$=1.06 which hosts a radio-loud quasar \citep{Siemiginowskaetal05, Siemiginowskaetal10}. The ICM properties are well constrained at large radii but in the central $\sim$30~kpc the X-ray bright AGN dominates the emission, making a detailed study of conditions in the nuclear region difficult. This is at the outer edge of the range of radii of maximum thermal instability in nearby clusters \citep{McCourtetal12}, where cooling and precipitation from the ICM are expected to be most effective. Cluster central quasars may also have an impact on the ICM, their intense radiation fields causing Compton cooling \citep{Russelletal10,Walkeretal14}, adding to the complexity of using such as system as a proxy for cooling in low-redshift, FR~I-dominated clusters. Another example, IRAS~F15307+3252, is a hyper-luminous infrared galaxy at $z$=0.93 which hosts an obscured quasar and CSS radio source \citep{HlavacekLarrondoetal17}. This object is hosted by a relatively X-ray faint galaxy group, for which only global properties could be determined.

A third known cluster-central CSS radio galaxy is 1321+045, located at $z$=0.263 in MaxBCG J201.08197+04.31863, a $\sim$4.4~keV cluster \citep{Kunert-Bajraszewskaetal13}. As well as being at lower redshift than the other two examples, 1321+045 is a relatively low-power radio source ($L_{5GHz}$$\sim$10$^{25}$~W~Hz$^{-1}$) without an X-ray bright quasar core. A \chandra\ snapshot observation in 2011 confirmed the presence of an extended ICM with a cool core. Analysis of the galaxy population suggests that the cluster is relatively relaxed \citep{WenHan13}, while the Sloan Digital Sky Survey (SDSS) spectrum of the brightest cluster galaxy (BCG) reveals strong H$\alpha$ emission with a luminosity comparable to that observed in rapidly cooling clusters at low-redshift \citep[L$_{\rm H\alpha}$=4.5$\times$10$^{41}$\ergps,][]{Liuetal12}. 

In this paper, we use a combination of new Very Long Baseline Array (VLBA) radio, \chandra\ X-ray, and IRAM 30m telescope CO observations, plus archival Very Large Array (VLA) and Multi-Element Radio Linked Interferometer Network (MERLIN) radio data to investigate 1321+045. Our goal is to examine the current state of the CSS source, its host galaxy, and the surrounding galaxy cluster, with a view to determining whether it provides a good example of the earliest stages of AGN feedback in galaxy clusters. Throughout the paper we adopt a flat cosmology with H$_0$=70\kmpspMpc, $\Omega_\Lambda$=0.7 and $\Omega_{\rm M}$=0.3. We adopt a redshift for the BCG of $z$=0.263, which gives an angular scale of 1\arcs=4.058~kpc, luminosity distance 1335.1~Mpc and angular distance 836.9~Mpc.


\section{Radio observations --- continuum emission}
\label{sec:vlba}
\subsection{Instruments and data reduction}

\begin{figure*}[ht]
\centering
\includegraphics[width=7cm,height=9cm]{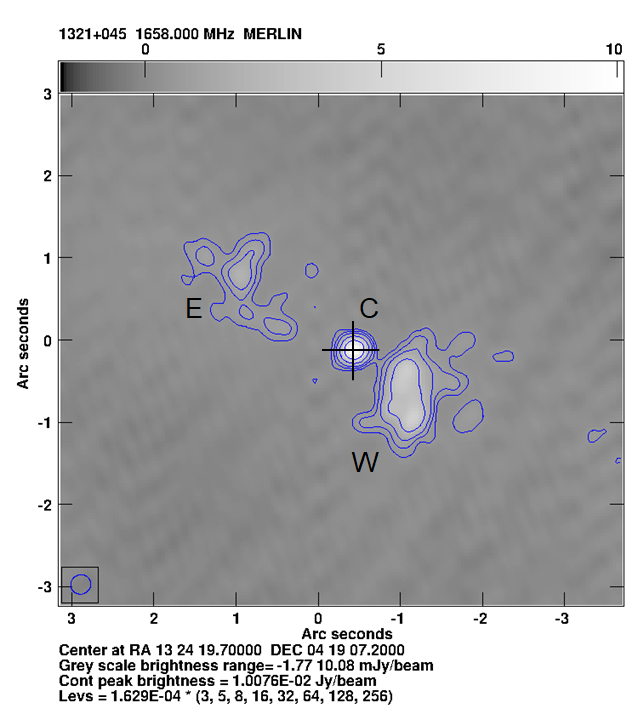}
\includegraphics[width=7cm,height=9cm]{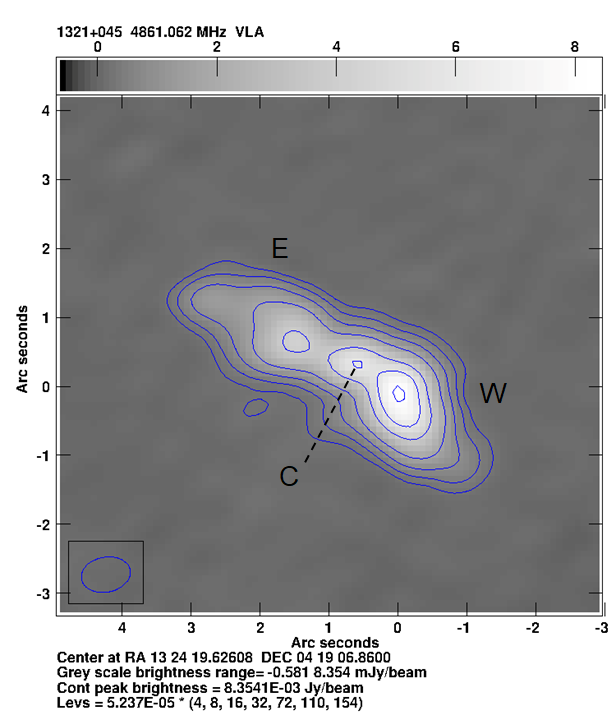}\\
\includegraphics[width=7cm,height=9cm]{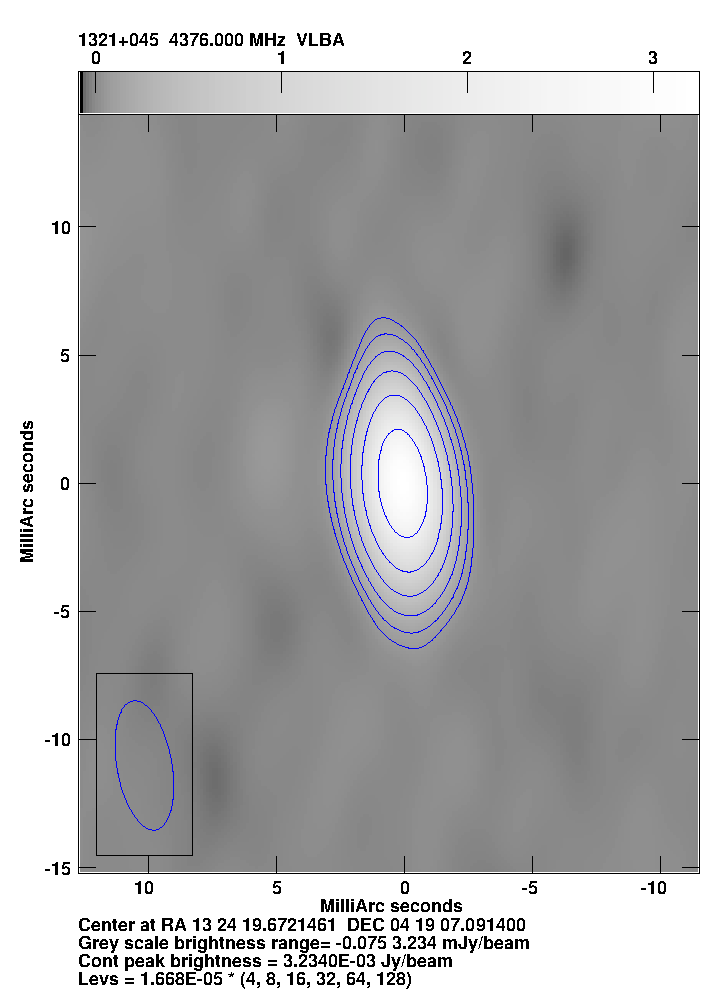}
\includegraphics[width=7cm,height=9cm]{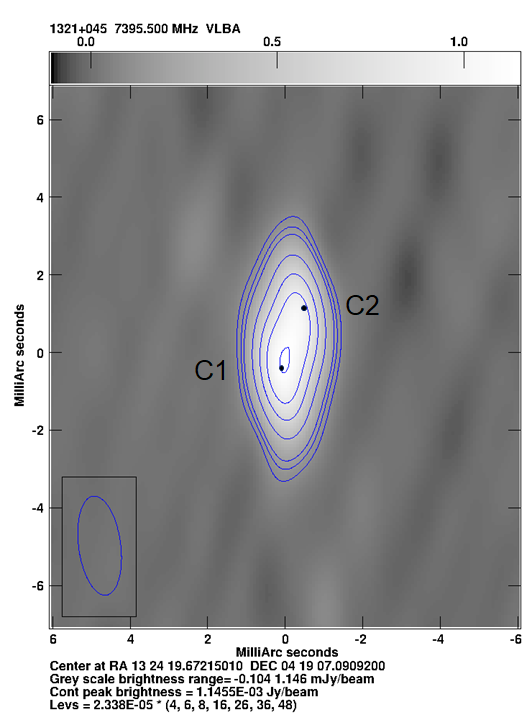}
\caption{{\bf Top}: MERLIN 1.6\,GHz and VLA 5\,GHz images of galaxy 1321$+$045. The radio lobes (E and W) and the core (C) are marked. The black cross indicates the position of SDSS optical counterpart. The beam size and r.m.s. noise levels of the MERLIN and VLA images are 0.25\arcs$\times$0.24\arcs, 163~$\mu$Jy, and 0.72\arcs$\times$0.51\arcs, 52~$\mu$Jy respectively. {\bf Bottom}: The high-resolution VLBA 4.5 and 7.5\,GHz images of radio core (C) of 1321$+$045. At 7.5 GHz the radio core splits into two components C1 and C2. The beam size and r.m.s. noise levels of the 4.5 and 7.5~GHz images are 4.09mas$\times$1.73mas, 17~$\mu$Jy, and 2.58mas$\times$1.03mas, 23~$\mu$Jy respectively.}
\label{fig:Merlin_VLBA}
\end{figure*}

The radio source 1321$+$045 (R.A.~=13$^{\rm h}$24$^{\rm m}$19$^{\rm s}$.7,
Decl.~=~+04$\degree$19$\arcmin$07.2$\arcsec$, J2000.0) belongs to the class
of young CSS radio sources. It has been observed with MERLIN at 1.6\,GHz in
2006 December as a part of a large sample of low luminosity compact (LLC)
sources \citep{Kunert-Bajraszewskaetal10}.  The data reduction was made
using an AIPS-based PIPELINE procedure developed at Jodrell Bank
Observatory (JBO) and the NRAO AIPS\footnote{http://www.aips.nrao.edu}
software. The 1.6\,GHz MERLIN image is reproduced here in Figure
\ref{fig:Merlin_VLBA}.

\begin{table*}
\caption{\label{tab:radio}Sensitivity and beam parameters for the radio continuum observations. Position angles are measured anti-clockwise from north.}
\begin{center}
\begin{tabular}{lcccccc}
\hline
\hline
Observatory & project & observation date & frequency & beam size & p.a. & r.m.s. noise \\
 & & & (GHz) & & (\degree) & ($\mu$Jy) \\
\hline
MERLIN & MN-R-06B-013 & 2006 Dec 12 & 1.66 & 0.25\arcs$\times$0.24\arcs & -62.95 & 163 \\
VLA    & AK0360       & 1994 May 07 & 4.86 & 0.72\arcs$\times$0.51\arcs & -79.05 & 52 \\
VLBA   & BK225        & 2020 Mar 18 & 4.5  & 4.09mas$\times$1.73mas & 4.63 & 17 \\
       &              &             & 7.5  & 2.58mas$\times$1.03mas & 7.34 & 23 \\
\hline
\end{tabular}
\end{center}
\end{table*}

In addition we have obtained a 5\,GHz image of 1321$+$045 using archival VLA data (Figure \ref{fig:Merlin_VLBA}). It is a short $\sim$1.5 min scan on this object made in 1994 May (PI: Kollgaard) in the BnA configuration. The data reduction was performed with AIPS software and source 3C286 was used as an amplitude and phase calibrator. The flux densities of the main components of the source in MERLIN and VLA images were then measured using AIPS tasks JMFIT or IMEAN.

New high spatial resolution observations of 1321+045 were made with the VLBA at C-band on 2020 March 18 in the phase-referenced mode. The data were correlated at the Array Operations Center in Socorro (USA). In order to carry out the spectral analysis of the radio components of our source we divided the available bandwidth
of the C-band receiver into two sub-bands centered
at 4.5 and 7.5 GHz. Data reduction (including editing, amplitude calibration, instrumental phase corrections and fringe-fitting) was performed following the standard
procedure using the NRAO AIPS software. After this stage the AIPS task IMAGR was used to produce the final total intensity images. The flux densities of the main components of the source were then measured by fitting
Gaussian models using AIPS task JMFIT. The linear size of the source was calculated based on the largest angular size measured in the 7.5 GHz image contour plot. Table~\ref{tab:radio} summarizes the sensitivities and spatial resolutions of the various radio continuum datasets.

\subsection{Results}
\label{sec:vlba_res}
The arcsecond-scale radio morphology of 1321$+$045 is shown on MERLIN 1.6\,GHz and VLA 5\,GHz images at resolution $0.3\times0.2$ and $0.7\times0.5$ arcseconds, respectively (Figure~\ref{fig:Merlin_VLBA}). The structure consists of three components well separated on the higher resolution MERLIN image. The position of the central component (marked C) is well correlated with the SDSS position of the optical counterpart (black cross) suggesting it is a radio core. 
The two lobes (E and W) are located on the opposite sides of the core. There is no evidence of jets or hotspots that would connect the radio lobes with nucleus C. The flux densities of the separated components E, C and W amount to $38\pm4$, $11\pm2$ and $49\pm5$ mJy in the MERLIN map, and $12\pm2$, $4\pm1$ and $20\pm2$ mJy in the VLA map, respectively. However, the MERLIN observations show $\sim 15-20 \%$ loss of the total flux density at 1.6 GHz probably connected with
the extended emission of the radio lobes. Taking this into account we can give only a rough estimation of the spectral index\footnote{We define spectral index $\alpha$ as S$_\nu\propto\nu^{-\alpha}$ where S$_\nu$ is the flux density at frequency $\nu$.} of the components E, C and W that amounts to $\alpha^{5~GHz}_{1.6~GHz}$=1.1, 0.9 and 0.8, respectively. 


The total projected length of the source is $\sim 16\,$kpc and its radio luminosity, $L_{5{\rm GHz}} \sim 10^{25}$\,W~${\rm Hz^{-1}}$ ($< 10^{42}$\,erg~${\rm s^{-1}}$), places it in the FR\,I-FR\,II transition region.

Inspection of the VLSS image of 1321+045 at 74\,MHz \citep{Cohenetal07} shows that the low frequency emission is uniform and consistent with the expected beam; there is no trace of more extended components which could indicate an older phase of radio activity.

The new high-resolution ($\rm 5\times2 mas$) VLBA observations at 4.5 and 7.5 GHz show the radio core which splits into two components (C1 and C2) at the higher frequency (Figure \ref{fig:Merlin_VLBA}). The total flux density of the component C at 4.5 GHz amounts to $3.86\pm0.04$ mJy which is in agreement with the value measured for the central component in the VLA 5\,GHz image at arcsec resolution.
At 7.5 GHz the total flux densities of the components C1 and C2 amount to $1.24\pm0.03$ and $0.72\pm0.04$ mJy, respectively. We interpret the feature C2 as a radio jet, with a total projected linear size $\sim$20~pc. Its orientation, however, is different to that of the arcsecond-scale MERLIN radio lobes, which raises the possibility that it represents a new phase of radio activity which has recently begun, with the outer lobes of 1321+045 no longer supplied with fresh particles and evolving passively in the coasting phase.

Summing the 7.5~GHz fluxes of components C1 and C2, we can estimate the overall spectral index of the core and small-scale jet to be $\alpha^{7.5~GHz}_{4.5~GHz}$=1.33$\pm$0.07. The spectral index estimated for the core+jet from the MERLIN and VLA data (0.9) is a little flatter, but may not be reliable as discussed above. These values are quite steep, and in principle newly launched jets might be expected to have a relatively flat spectrum ($\alpha$$\sim$0.5). However, young, small-scale GPS and CSS radio sources often have steep spectral indices ($\sim$1) above their spectral break \citep[e.g.,][]{Wolowskaetal21}. The core and new jet in 1321+045 may be an example of these classes of source, but additional high-resolution observations would be needed to measure their spectra.

\section{Radio observations --- line emission}
1321+045 was observed by the IRAM 30m telescope on 2019 December 8-10 (project 208-19), in good weather conditions. The EMIR receivers were used to observe the redshifted CO(1-0) and CO(3-2) lines simultaneously, at 91.263~GHz and 273.803~GHz respectively, using the E0 and E2 bands and both horizontal and vertical polarisations. The CO(2-1) line could not be observed as it is redshifted into an atmospheric absorption band. The FTS and WILMA backends were used at both frequencies, and the source was observed for a total of $\sim$12 hours. Observations of bright quasars or planets were used to calibrate the pointing and focus of the telescope. The data were reduced using the \textsc{gildas}\footnote{http://www.iram.fr/IRAMFR/GILDAS} \textsc{class} software \citep{Pety05,GILDAS13}.

\begin{figure}
\includegraphics[width=8.4cm,viewport=20 50 680 400,clip=true]{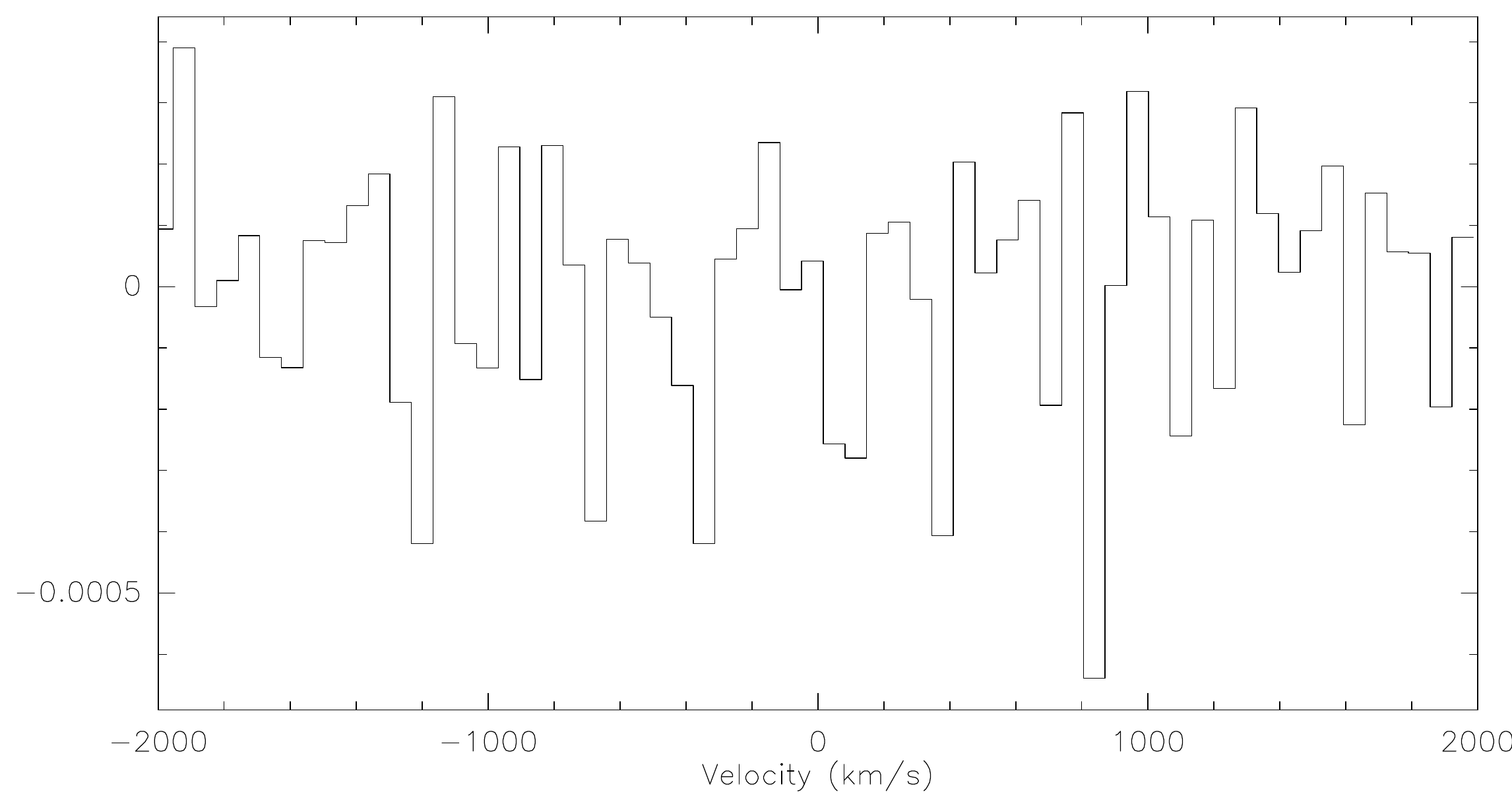}
\includegraphics[width=8.4cm,viewport=20 50 680 400,clip=true]{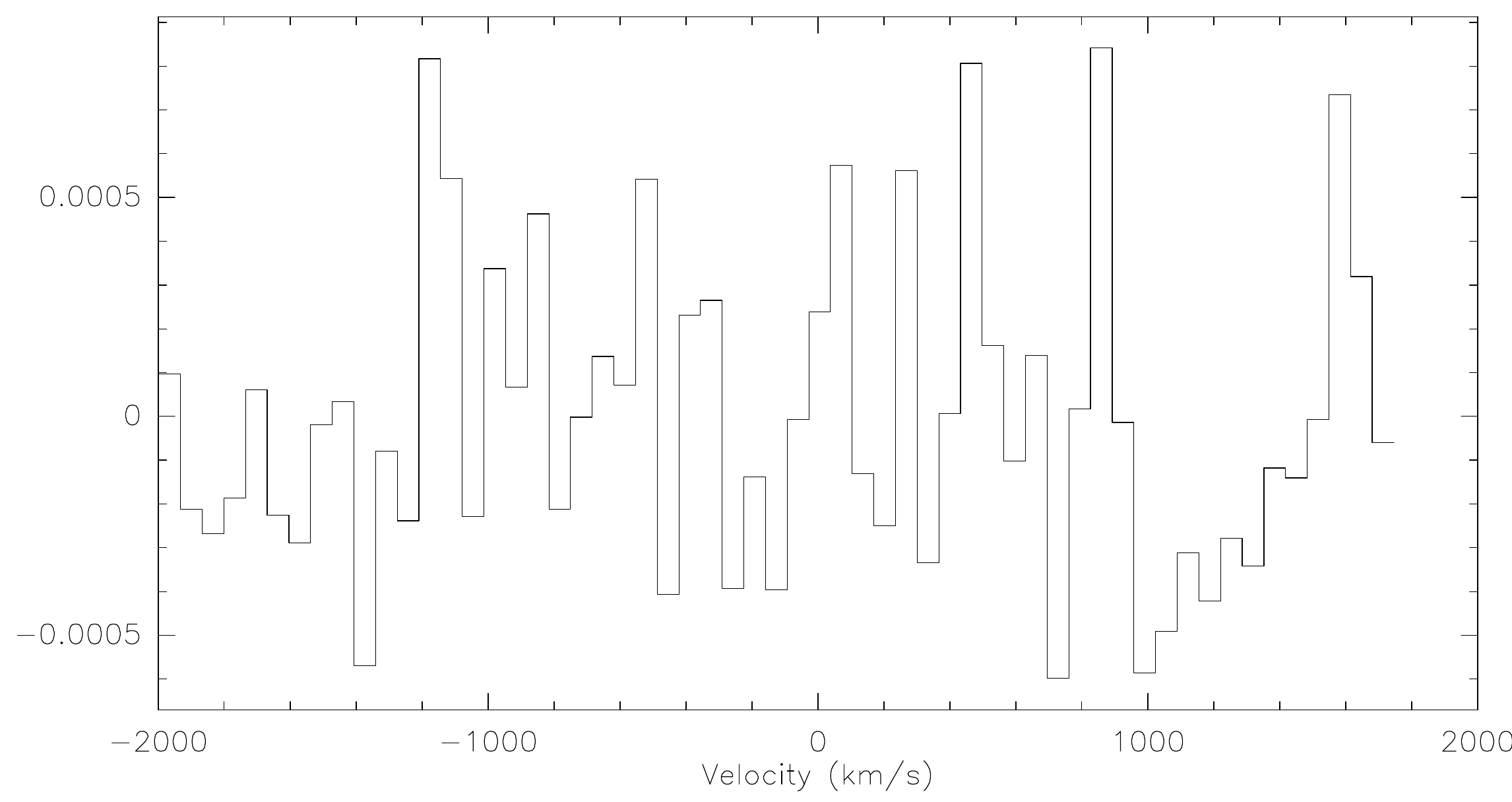}
\caption{\label{fig:CO}IRAM 30m spectra centred on the redshifted CO(1-0) and CO(3-2) lines, in the upper and lower panels respectively. The vertical axes indicate antenna temperature T$_{\rm a}^*$ in K.}
\end{figure}

Figure~\ref{fig:CO} shows the final summed CO(1-0) and CO(3-2) spectra for 1321+045. Neither line was detected, but the data are deep enough to place strong upper limits on the source. Binned to 65\kmps, the r.m.s. antenna temperature T$_{\rm a}^*$ was 0.23~mK for the CO(1-0) spectrum and 0.75~mK for the CO(3-2) spectrum.

We follow the approach described in \citet{OSullivanetal18}, adopting a nominal linewidth of 300\kmps\ and CO-to-H$_2$ conversion factor $\alpha_{CO}$=4.6 \citep[including Helium,][]{SolomonVandenbout05}. We also apply a correction factor of R$_{13}$=L$^{\prime}_{\rm CO 1-0}$/L$^{\prime}_{\rm CO 3-2}$=2 when converting the CO(3-2) luminosity to a mass, to account for the lower Rayleigh-Jeans brightness temperature of the (3-2) transition \citep{Tacconietal13}. We calculate the 3$\sigma$ upper limits on the molecular gas mass to be M$_{\rm mol}\leq$7.7$\times$10$^9$\Msol\ from the CO(1-0) data, and $\leq$5.6$\times$10$^9$\Msol\ from the CO(3-2), assuming the CO(3-2) line to be fully excited. This is a factor of 1.3-2.8 below the expected mass of $\sim$10$^{10}$\Msol\ predicted from the L$_{\rm H\alpha}$:M$_{\rm mol}$ relation \citep{SalomeCombes03}.

\section{Chandra X-ray Observations and Data Reduction}
\label{sec:obs}

\begin{table*}
\caption{\label{tab:obs}Summary of the \chandra\ ACIS-S observations.}
\begin{center}
\begin{tabular}{lcccc}
\hline
ObsID & PI & Observation & Observing & Cleaned \\
 & & date & mode & exposure (ks) \\
\hline
12715 & Kunert-Bajraszewska & 2011 Dec 14 & FAINT, 1/8 subarray & 9.3 \\
22634 & O'Sullivan & 2020 Mar 31 & VFAINT & 29.7 \\
23068 & O'Sullivan & 2020 Apr 20 & VFAINT & 28.7 \\
23204 & O'Sullivan & 2020 Apr 05 & VFAINT & 25.6 \\
\multicolumn{4}{r}{Total:} & 93.3 \\
\hline
\end{tabular}
\end{center}
\end{table*}

1321+045 was observed by \chandra\ during cycles~12 and 21, using the ACIS-S detector. Table~\ref{tab:obs} provides some details of the observations. A summary of the \chandra\ mission is provided in \citet{Weisskopfetal02}. The cycle~12 observation was performed in 1/8 subarray mode, as a precaution against pile-up. This was not necessary for the later observations, as the AGN was found to be relatively X-ray faint. We processed the observations using \textsc{ciao}~4.12 \citep{Fruscioneetal06} and CALDB~4.9.1, following the approach laid out in the \chandra\ analysis threads\footnote{https://cxc.harvard.edu/ciao/threads/index.html} and \citet{OSullivanetal17}.

Each observation was filtered for periods of high background using the \textsc{lc\_clean} script. We used the standard set of \chandra\ blank-sky background files to create background spectra and images. The background was normalised to each observation using the 9.5-12~keV count rate. Very faint mode filtering was applied to the cycle~21 observations and background files. 

All observations were reprojected to a common tangent point, and combined images and exposure maps were created using the \textsc{reproject\_obs} and \textsc{merge\_obs} tasks. We used the combined 0.5-7~keV image to identify point sources in the field of view, via the \textsc{wavdetect} task. Examination of the point sources showed that there were no significant astrometric differences between observations. Point source extent was determined based on a combined map of the telescope point spread function (PSF), and chosen to ensure the exclusion of at least 90 per cent of the point source flux. Point sources were excluded from all further analysis. An apparent source was found coincident with the peak of cluster diffuse emission, but examination showed it to be a false detection. 

The overall astrometry of the X-ray data was tested by searching for optical counterparts to the X-ray point sources. Of seven sources with counterparts within 3.5\arcm\ of the BCG, four have positions matching to within $<$0.5\arcs, and a fifth matches within 0.66\arcs. This is consistent with the expected astrometric accuracy of \chandra. The other two sources have greater offsets, but their extent is large enough ($>$5\arcs radius) that the apparent counterparts may be chance correlations. For the five sources with well-matched counterparts, we see no trend in the direction of the position offset from the optical, so we conclude that the \chandra\ astrometry is accurate and reliable enough for our purposes.

Image analysis was performed using combined images from all four exposures. Spectra were extracted separately for each observation, using the \textsc{specextract} task and, for the cycle~21 observations, combined into single sets of spectra and responses for each region using \textsc{combine\_spectra}. Initial examination of spectra showed the blank-sky background over-estimated the soft background component. We therefore applied a correction to each background spectrum, based on the scaled difference between the observation and blank-sky background spectra extracted from a source-free region of the S1 CCD (for the cycle~21 observations) or S3 (for ObsID 12715, where S1 was not used). We fitted spectra using either \textsc{Xspec}~12.11.0h \citep{Arnaud96} or \textsc{ciao sherpa} \citep{Freemanetal01}. We adopted a Galactic hydrogen column of 2.12$\times$10$^{20}$\pcmsq, which includes the contribution from molecular hydrogen \citep{Willingaleetal13}, and the solar abundance ratios of \citet{GrevesseSauval98}.

\section{chandra X-ray Results}
\label{sec:res}

\begin{figure*}
\includegraphics[width=8.4cm,bb=36 126 577 667]{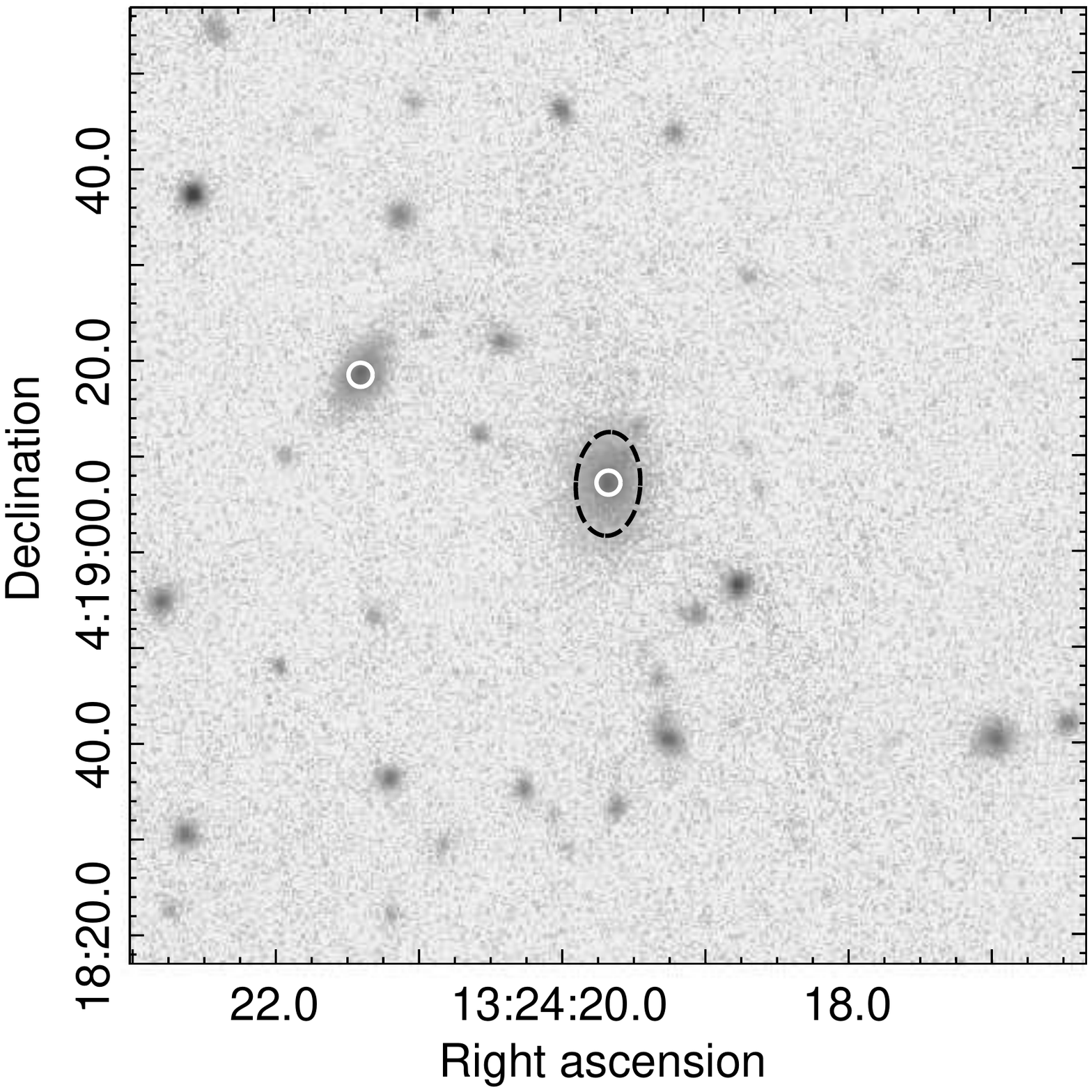}
\includegraphics[width=8.4cm,bb=36 126 577 667]{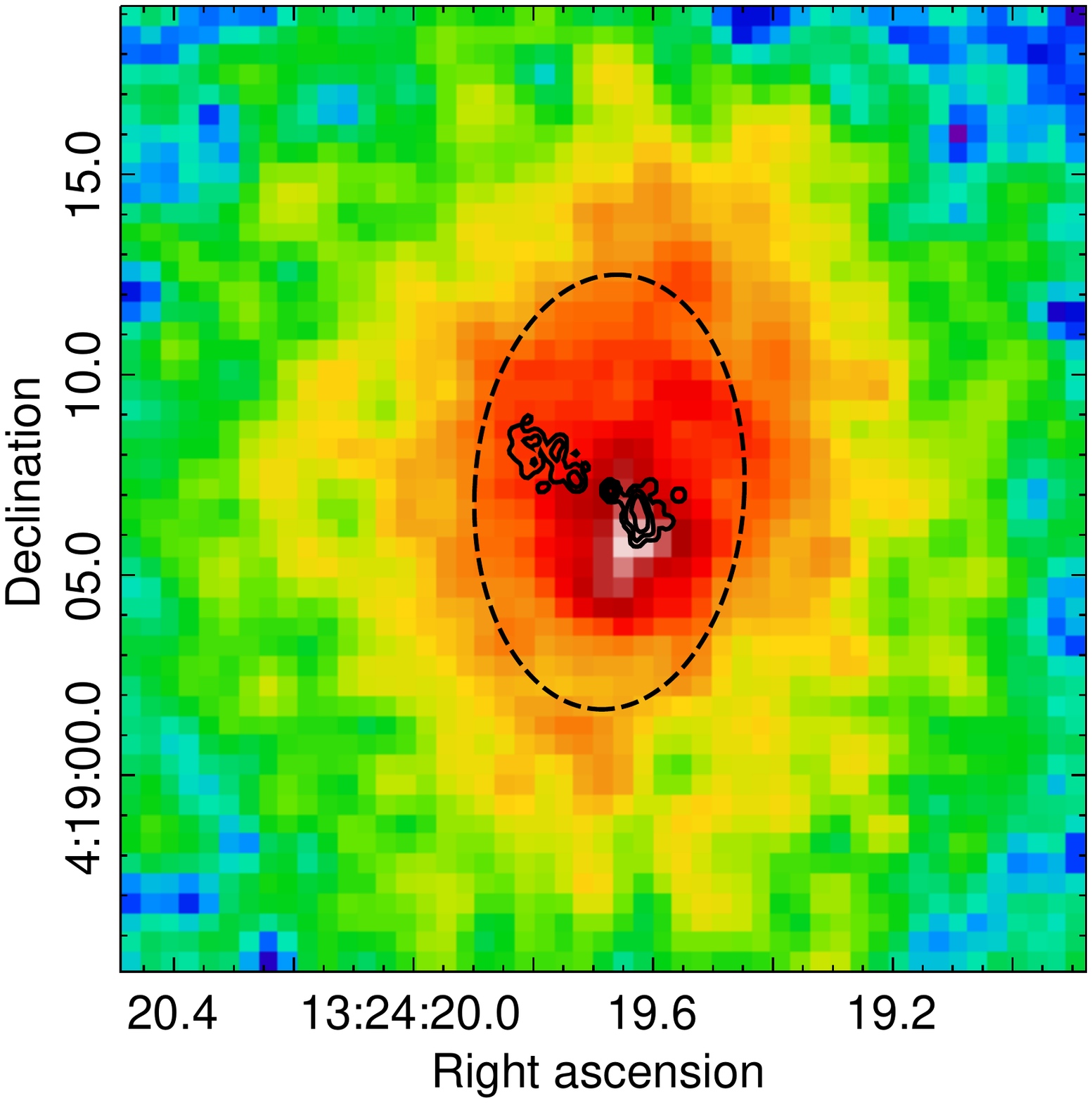}
\caption{\label{fig:im}Images of the core of the 1321+045 cluster. The left panel shows a PanSTARRS $r$-band image of the central $\sim$200~kpc ($\sim$50\arcs) of the cluster with the two largest member galaxies marked by white circles. The dashed ellipse shows the approximate \Dtf\ contour of the BCG. The right panel shows an exposure corrected \chandra\ 0.5-7~keV image of the central $\sim$50~kpc, smoothed with a 1\arcs\ Gaussian, with MERLIN 1.6~GHz contours overlaid, and the \Dtf\ ellipse marked. The radio core is located at the optical centroid of the BCG.}
\end{figure*}

\chandra\ and optical images of 1321+045 are shown in Figure~\ref{fig:im}. The radio core is located at the optical centre of the BCG, but falls $\sim$1.5\arcs/6~kpc to the north of the broad band (0.5-7~keV) X-ray peak. Examination of the soft and hard band (0.5-2 and 2-7~keV) X-ray images show that the X-ray peak in the broad band image is driven by a peak in the soft band emission just to the south of the radio core. The large-scale diffuse hard band emission from the ICM is fairly well-centred on the BCG nucleus, but there is no indication of a small-scale hard-band peak or point source associated with the AGN.

To understand the structure in the X-ray images better, we modelled the 0.5-7~keV surface brightness distribution using \textsc{ciao sherpa}. As the difference in location between the hard and soft band peaks suggests, the image is not well described by a simple $\beta$-model. We used a model consisting of 2 elliptical $\beta$-model components and a circular Gaussian, plus a flat background component, all folded through the combined exposure map. Table~\ref{tab:SB} shows the fit parameters. The Gaussian is relatively compact and corresponds to the (broad band and soft band) X-ray peak. The more extended $\beta$-model describes the large-scale ICM emission. The second $\beta$-model has an unphysically steep outer slope, but describes some excess emission above the larger ICM model, centred north of the X-ray peak and aligned NNW. The major axis of the more extended ICM component is aligned just east of north.

\begin{table}
\caption{\label{tab:SB}Best fitting surface brightness model parameters. Position angles are measured anti-clockwise from west.}
\begin{center}
\begin{tabular}{lcc}
\hline
component & parameter & best fit \\
\hline
Gaussian & FWHM & 2.54$\pm$0.26\arcs\ \\
 & R.A. & 13 24 19.64 \\
 & Dec. & +04 19 05.65 \\
$\beta$-model 1 & r$_{\rm core}$ & 16.6$^{+1.6}_{-0.98}$\arcs\\
 & $\beta$ & 0.59$\pm$0.01 \\
 & ellipticity & 0.30$\pm$0.01 \\
 & P.A. & 97.6$\pm$0.9\degree\\
 & R.A. & 13 24 18.72 \\
 & Dec. & +04 19 05.52 \\
$\beta$-model 2 & r$_{\rm core}$ & 16.1$^{+3.1}_{-4.2}$\arcs\\
 & $\beta$ & 2.8$^{+4.8}_{-1.1}$\\
 & ellipticity & 0.12$\pm$0.04\\
 & P.A. & 65.2$\pm$10.9\degree\\
 & R.A. & 13 24 19.63\\
 & Dec. & +04 19 08.06\\
\hline
\end{tabular}
\end{center}
\end{table}

\begin{figure}
\includegraphics[width=8.4cm,bb=36 126 577 667]{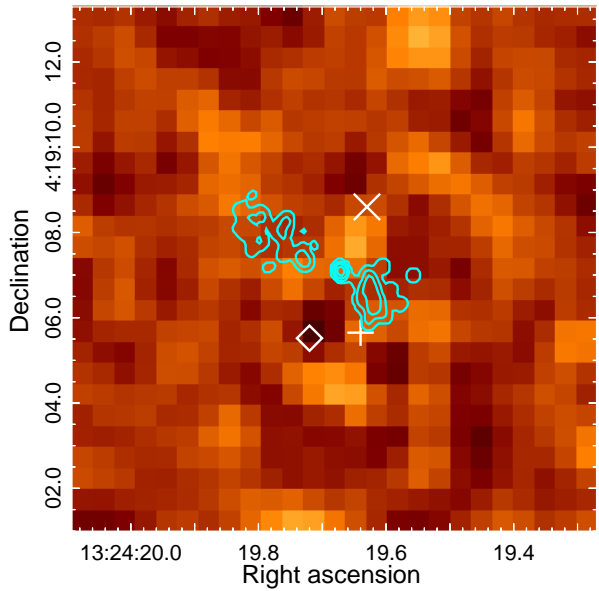}
\caption{\label{fig:resid}\chandra\ 0.5-7~keV residual map of the cluster core after subtraction of the best-fitting surface brightness model. The image is lightly smoothed with a 1\arcs\ Gaussian. MERLIN 1.6~GHz contours are overlaid, and the cross, diamond, and X symbols mark the centroid positions of the Gaussian, first and second $\beta$-model components respectively.}
\end{figure}

Figure~\ref{fig:resid} shows the 0.5-7~keV residual image of the cluster core, produced by subtracting the best-fitting surface brightness model. Comparison with the radio emission again shows the AGN core to be slightly to the north of the peak of X-ray emission. The centroids of the Gaussian and more extended $\beta$-model surface brightness components are close together, south of the AGN, whereas the second $\beta$-model is centred slightly to the north. The residual map does not show any clear structures correlated with the radio source. 

\begin{figure}
\centering
\includegraphics[width=8.4cm,bb=36 80 577 710]{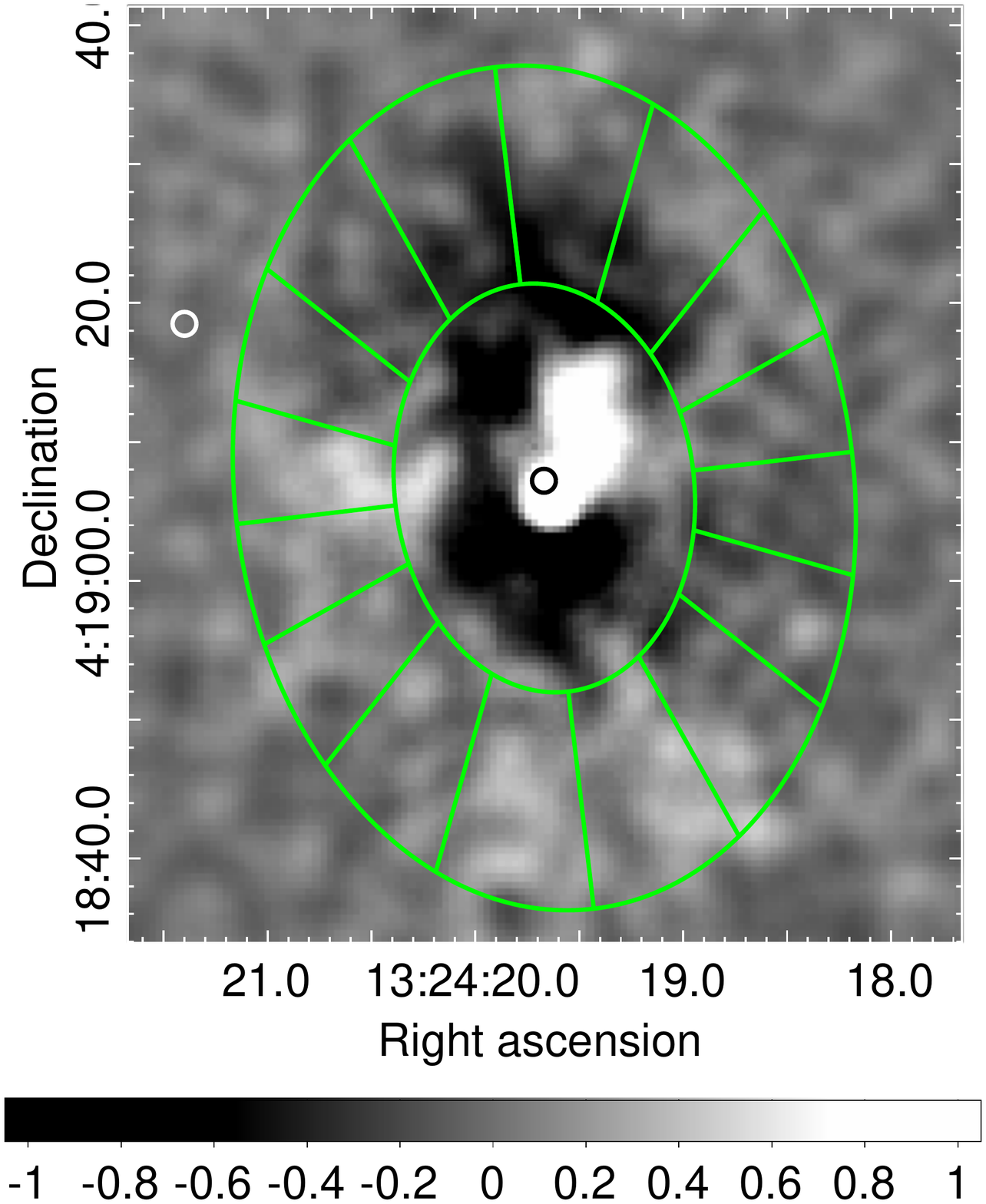}
\includegraphics[width=8.4cm,bb=30 350 555 730]{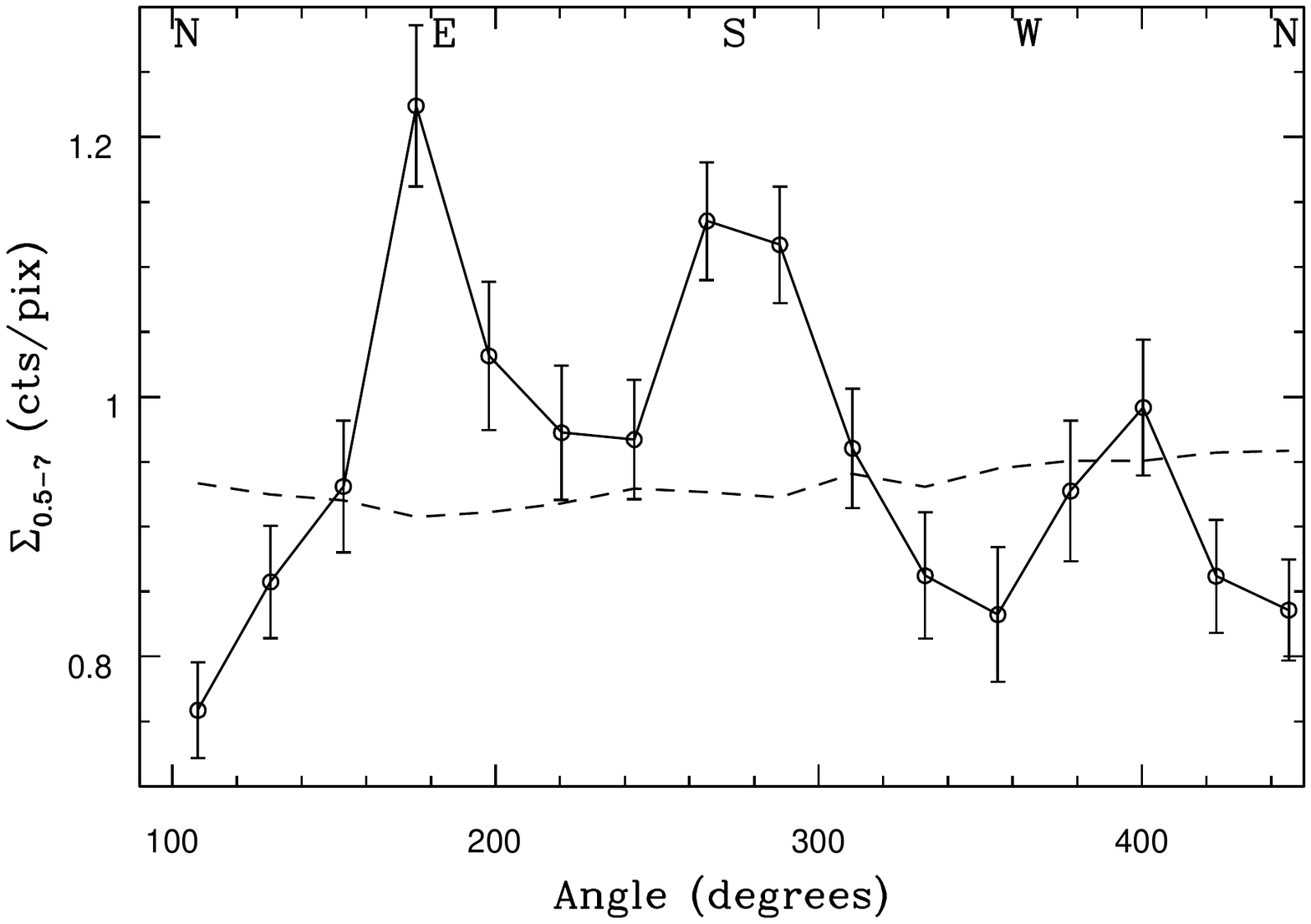}
\caption{\label{fig:azimuthal}\chandra\ 0.5-7~keV residual map, smoothed with a 2.5\arcs\ Gaussian, after removal of a single $\beta$-model describing the large-scale ICM emission. Brighter and darker regions represent positive and negative residuals to the model. The two circles mark the positions of the BCG and the second brightest galaxy. The lower panel shows the azimuthal variation in surface brightness in the $\sim$15-30\arcs\ elliptical annulus marked on the image. Error bars indicate 1$\sigma$ uncertainties, and the dashed line the expected surface brightness from the $\beta$-model. Note that the colour bar on the upper panel shows the scale for \textit{smoothed} counts, whereas the lower panel shows the true measured values from the \textit{unsmoothed} data.}
\end{figure}

Figure~\ref{fig:azimuthal} shows a residual map on a larger scale, with only the large-scale $\beta$-model component representing the ICM subtracted. The emission which would be modelled by the other two model components is visible as a bright excess at the location of the BCG and extending in a tail to the NNW. At larger radii, $\sim$15-30\arcs\ ($\sim$60-120~kpc) there is an excess above the model to the south and east, and oversubtraction to the west and particularly the north. The southern excess corresponds to the base of a cool tail visible in the spectral maps (see Section~\ref{sec:maps}). The brightest part of the excess to the east also corresponds to a region of cooler emission, and could suggest an arc of emission pointing roughly in the direction of the second brightest galaxy in the cluster. Using segments of an elliptical annulus to measure the surface brightness, we find that these excesses are 4-5$\sigma$ significant.

\subsection{Spectral mapping}
\label{sec:maps}
Spectral maps of the cluster were created using the ``fixed grid'' technique described in \citet{OSullivanetal19}. The field of interest was covered by a regular grid of 2\arcs\ map pixels, and spectra extracted from circular regions centered on each map pixel, with radius chosen to obtain a minimum of 2000 net counts. The resulting radii range from 4\arcs\ to 29.5\arcs, producing an effect analogous to adaptive smoothing across the map, with least smoothing in the nucleus and most in the outskirts. The combined spectrum for each region was then fitted using a \textsc{phabs*apec} absorbed thermal plasma model, with the absorption column fixed at the Galactic value. Temperature and abundance are measured directly from the fit, and pseudo-pressure and -entropy estimated from the fit parameters.

\begin{figure*}
\includegraphics[width=9.2cm,bb=36 51 577 742]{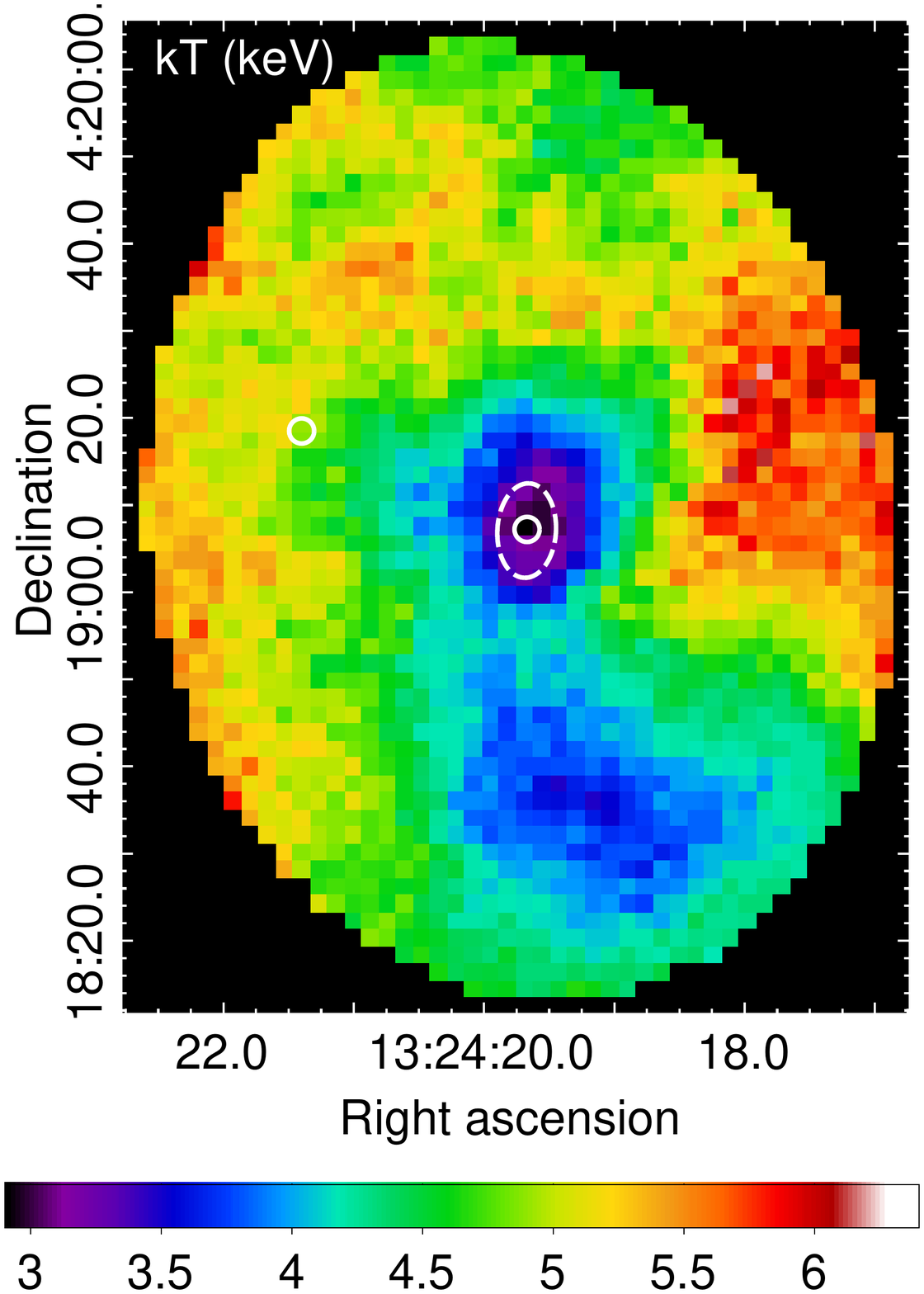}
\includegraphics[width=9.2cm,bb=36 51 577 742]{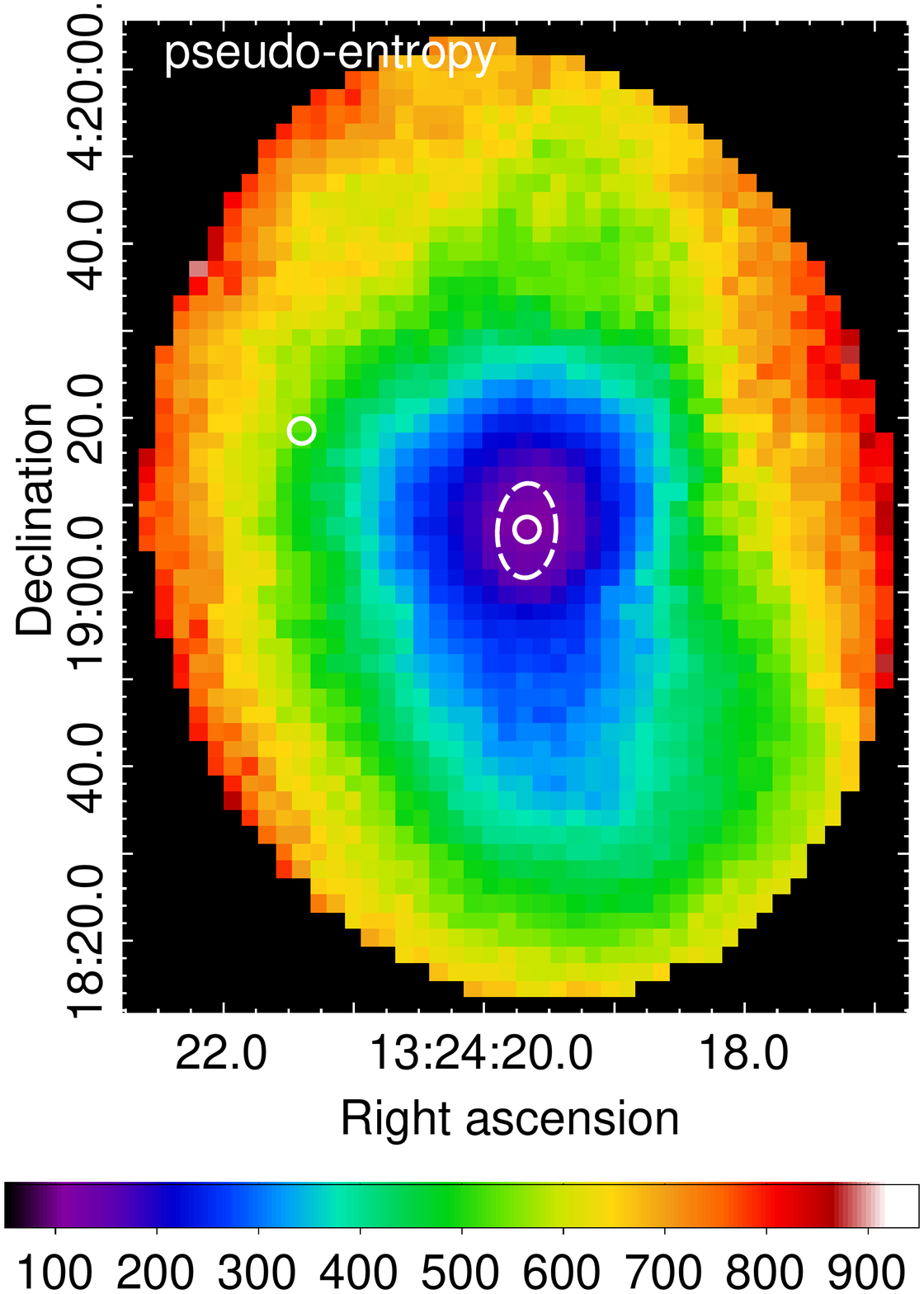}
\caption{\label{fig:maps}Spectral maps of the core of 1321+045. The left panel shows the temperature distribution (in keV) while the right panel shows pseudo-entropy. Note that the entropy scale is arbitrary, and the map should only be taken as an indication of the entropy distribution. Both maps have the same angular scale and orientation; pixels are $\sim$1\arcs\ across. The white circles indicate the locations of the two brightest galaxies in the cluster, and the \Dtf\ contour of the BCG is marked by a dashed ellipse.}
\end{figure*}

Figure~\ref{fig:maps} shows the temperature and pseudo-entropy maps. The temperature map shows a 3-3.5~keV cool core with an apparent $\sim$4~keV cool tail extending to the south. A small spur of cooler temperatures also extends to the east, along a line between the two brightest galaxies in the cluster, corresponding to the eastern surface brightness excess seen in Figure~\ref{fig:azimuthal}. Higher temperatures (5-6~keV) are found to the north and west. Typical 1$\sigma$ uncertainties on the temperature are $\sim$6\% in the cool core and $\sim$11\% in the hottest regions. The abundance map is relatively flat, probably indicating that the 2000 net count spectra do not have sufficient flux in the 6.7~keV Fe line to allow accurate abundance measurements. The pseudo pressure map is also relatively regular, showing a fairly circular distribution centred on the BCG. The pseudo-entropy map (Figure~\ref{fig:maps}) shows a strong N-S axis of lower entropies, and suggests that the cool tail consists of lower entropy material.

\subsection{Radial profiles of ICM properties}
\begin{figure*}
\includegraphics[width=\textwidth,bb=30 220 560 760]{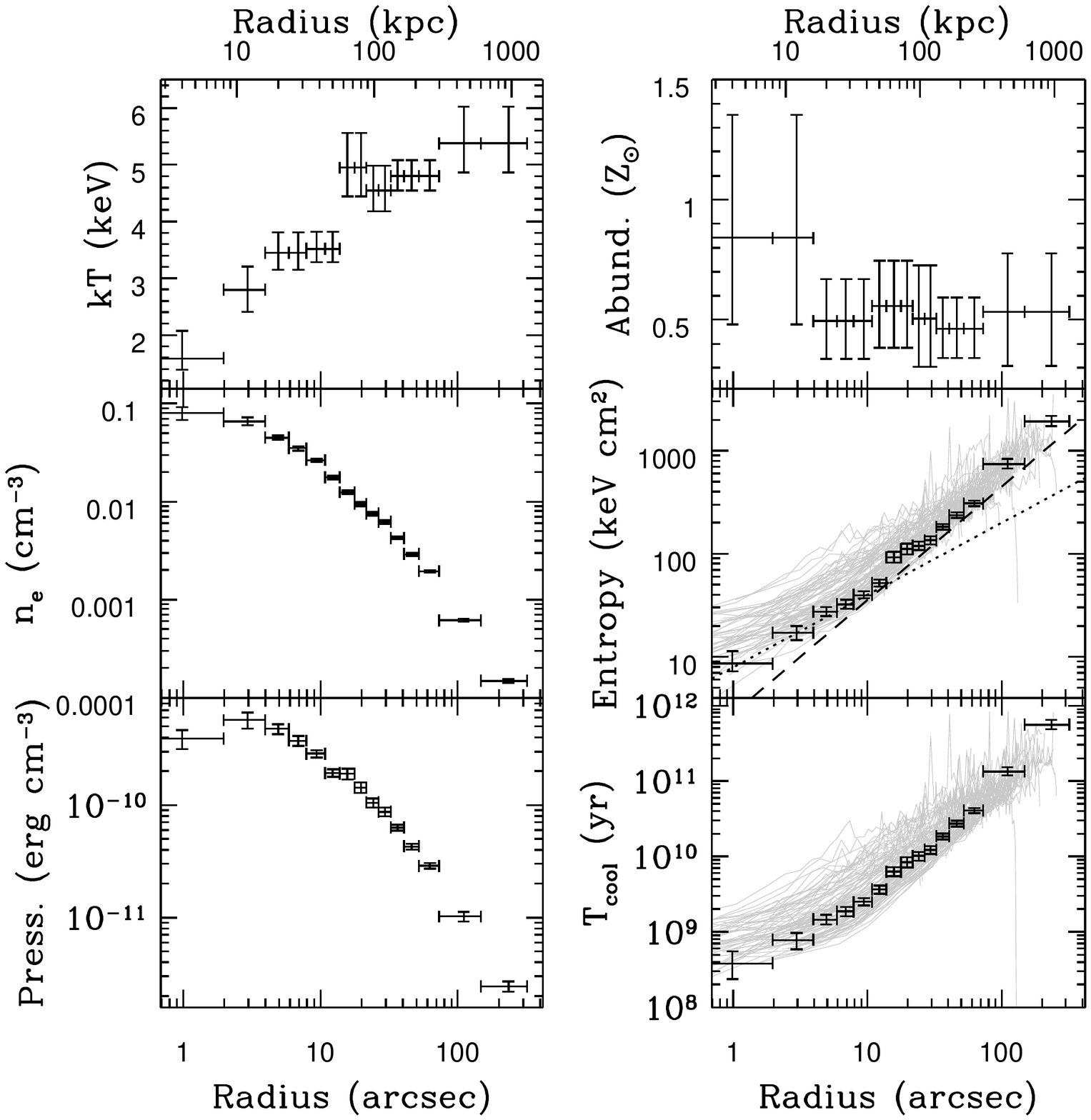}
\caption{\label{fig:deproj}Deprojected radial profiles of ICM properties. The dotted and dashed lines on the entropy plot show gradients of R$^{0.7}$ and R$^{1.1}$ respectively, with arbitrary normalizations. Grey lines on the entropy and isobaric cooling time plots show profiles from nearby strong cool core clusters of the ACCEPT sample \citep{Cavagnoloetal09}.}
\end{figure*}

We extracted spectra in a series of 13 circular annuli centred on the AGN and chosen to achieve signal-to-noise ratios of 40 or better in each annulus. The central region has a radius of $\sim$2\arcs\ ($\sim$8~kpc) and contains all three X-ray surface brightness centroids, the radio core, west lobe, and roughly half the east lobe. The spectra were fitted using a deprojected absorbed thermal plasma model \textsc{projct*phabs*apec} with the absorption set to the Galactic column. Pressure, entropy and cooling time were calculated from the temperature and electron density (n$_{\rm e}$) profiles, with pressure defined as 1.92n$_{\rm e}$kT and entropy n$_{\rm e}^{-2/3}$kT. The resulting profiles are shown in Figure~\ref{fig:deproj}.

The temperature and abundance profiles were quite unstable, and we were forced to tie these parameters in groups of bins in order to obtain a better constrained fit. The temperature profile is somewhat irregular, with temperatures $\sim$5~keV outside 14\arcs, falling to 1.6$^{+0.5}_{-0.2}$~keV in the central bin. The abundance profile is relatively flat, with a value $\sim$0.5\Zsol, though the uncertainties are large enough that a gradient may be present. Abundance is poorly constrained in the central two bins, even though their values are tied. Entropy in the central 8~kpc is low, 8.6$^{+2.7}_{-1.4}$\kevcmsq, and cooling time is short, 3.9$^{+1.7}_{-1.5}\times$10$^8$~yr. Cooling time falls below 7.7~Gyr within $\sim$80~kpc, and below 3~Gyr within $\sim$45~kpc. 

The profiles are generally smooth, but a sharp rise in temperature, with corresponding features in entropy and pressure, is visible at $\sim$14\arcs\ radius. This is the size of the central cool core visible in the temperature map, and outside this radius the temperature distribution is asymmetric, with the cool tail to the south but much hotter emission to the west. The pressure profile is inverted in the central bin. This is probably an indicator that the assumptions made in the deprojection are inaccurate, most likely that the gas is not spherically symmetric in the innermost regions. The offset surface brightness peak supports this explanation. The alternative explanation, that some other gas component contributes additional pressure support in the central bin, seems improbable, since the radio emission (the most likely candidate) is relatively weak and the lobes small compared with the volume of the region.

For self-similarly scaling systems, the entropy profile is expected to follow an R$^{1.1}$ profile, and in the outer part of the cluster (outside $\sim$30\arcs\ or $\sim$120~kpc) our measured entropy profile follows this relationship fairly well (see dashed line on Figure~\ref{fig:deproj}). It has been observed that in the central $\sim$100~kpc of clusters the entropy profile tends to be flatter, with a slope of approximately R$^{0.7}$ \citep{Panagouliaetal14}. Again, our measured profile matches this expectation well (dotted line) at least within the 14\arcs\ (56~kpc) temperature jump. Between 56 and 120~kpc the profile is clearly flatter than R$^{1.1}$, and falls above the R$^{0.7}$ line, presumably because of the hotter region to the west.


\subsection{Non-detection of the AGN}
Examination of combined \chandra\ images in various bands does not reveal any clear point source coincident with the AGN radio core. A spectrum extracted from a circle of radius 4.9\arcs\ centred on the radio core has 2500 net counts, and is well fitted by an absorbed APEC plasma, with kT=3.04$\pm$0.13~keV, 0.58$^{+0.14}_{-0.13}$\Zsol\ abundance, and the absorbing column fixed at the Galactic value (reduced $\chi^2$=0.84 for 107 degrees of freedom). The parameters of the thermal model are consistent with those found from the radial deprojection analysis of the ICM. Adding a power-law component does not improve the fit significantly, and the best-fitting normalisation of the power-law is consistent with zero within 2$\sigma$ uncertainties. The 3$\sigma$ upper limit on the luminosity of the powerlaw component is L$_{0.5-7keV}\leq$3.3$\times$10$^{43}$\ergps. Extracting a spectrum from a smaller 1.5\arcs\ region, the thermal plasma model is less well constrained, leading to a poorer constraint on the powerlaw flux. 


Absorption by dust and gas close to the AGN could be a factor in reducing the detected X-ray flux, for example if we are viewing it through a dusty torus. In this case the intrinsic absorbing column could be quite high. To investigate this possibility we expanded the energy range in our fitting to 0.5-10~keV to provide as strong a constraint as is possible with the \chandra\ spectrum of the central 4.9\arcs-radius region. There are only $\sim$40 net counts above 6~keV in the spectrum, and the simple absorbed APEC model described above fits it well (reduced $\chi^2$=0.81 for 112 d.o.f.). 

We then fitted a model with a Galactic absorption component, APEC thermal model to describe the ICM, plus a powerlaw affected by an intrinsic absorber at the redshift of the BCG (\textsc{phabs*[apec+zphabs*power]}). When fitted, the powerlaw index is poorly constrained and unphysical. Freezing it at $\Gamma$=1.7, a typical value for AGN, and freezing the APEC temperature and abundance, we find that the intrinsic absorption column and powerlaw normalisation trade off against one another; either the luminosity of the powerlaw component must be very low, or the intrinsic column must be high. Neither parameter is well constrained, and the powerlaw normalisation is again consistent with zero within 2$\sigma$. 

We therefore conclude that, while we cannot rule out a heavily absorbed powerlaw component from the AGN, the \chandra\ data offer no positive evidence of any AGN emission in the X-ray.

\section{Discussion}
\label{sec:disc}

One of our main goals in observing 1321+045 was to determine the current state of the AGN and its impact on the surrounding cluster emission. As CSS radio sources are believed to be young, a cluster central CSS source might be an opportunity to observe the earliest phases of AGN feedback.

However, the VLBA observations suggest that the outer, small-scale lobes detected in the MERLIN data may be remnants of a short outburst, with a new phase of jet activity already beginning. The projected orientation of the inner VLBA jet is close to 90$\degree$ offset from that of the lobes. Even if both jet and lobe axes have components along the line of sight, strong bends in the jet would be required for the two to be connected.

\subsection{AGN timescales}
\label{sec:timescales}
CSS radio sources are believed to be $<$10$^5$~yr old. However, as the radio lobes in 1321+045 may be associated with a previous cycle of jet activity, their age is somewhat uncertain. Studies of radio-loud AGNs suggest that low luminosity radio sources may have a different distribution of lifetimes. They can be short-lived radio objects with timescales of $10^4$ - $10^5$ years \citep{ReynoldsBegelman97,Czernyetal09,Kunert-Bajraszewskaetal10,Hardcastleetal19}. 

One way to estimate the ages of the radio structures is by reference to typical expansion timescales. The linear size of the inner 7.5 GHz core-jet structure amounts to $l\sim$20 pc. If we adopt the average separation speed of the hot spots in compact radio sources $v_{sep}=0.19\pm0.11 c/h$ \citep{GirolettiPolatidis09} we can estimate the age of the outer and inner structures to be about $3\times10^4$ and 350 years, respectively, assuming that they lie close to the plane of the sky. We must note, however, that the value of the separation speed used in the calculations is too small for a rapidly evolving jet, which we probably see in our source. Therefore, the estimated age of the inner structure should be considered an upper limit.

\begin{figure}
\centering
\includegraphics[width=8.4cm,bb=10 50 530 530]{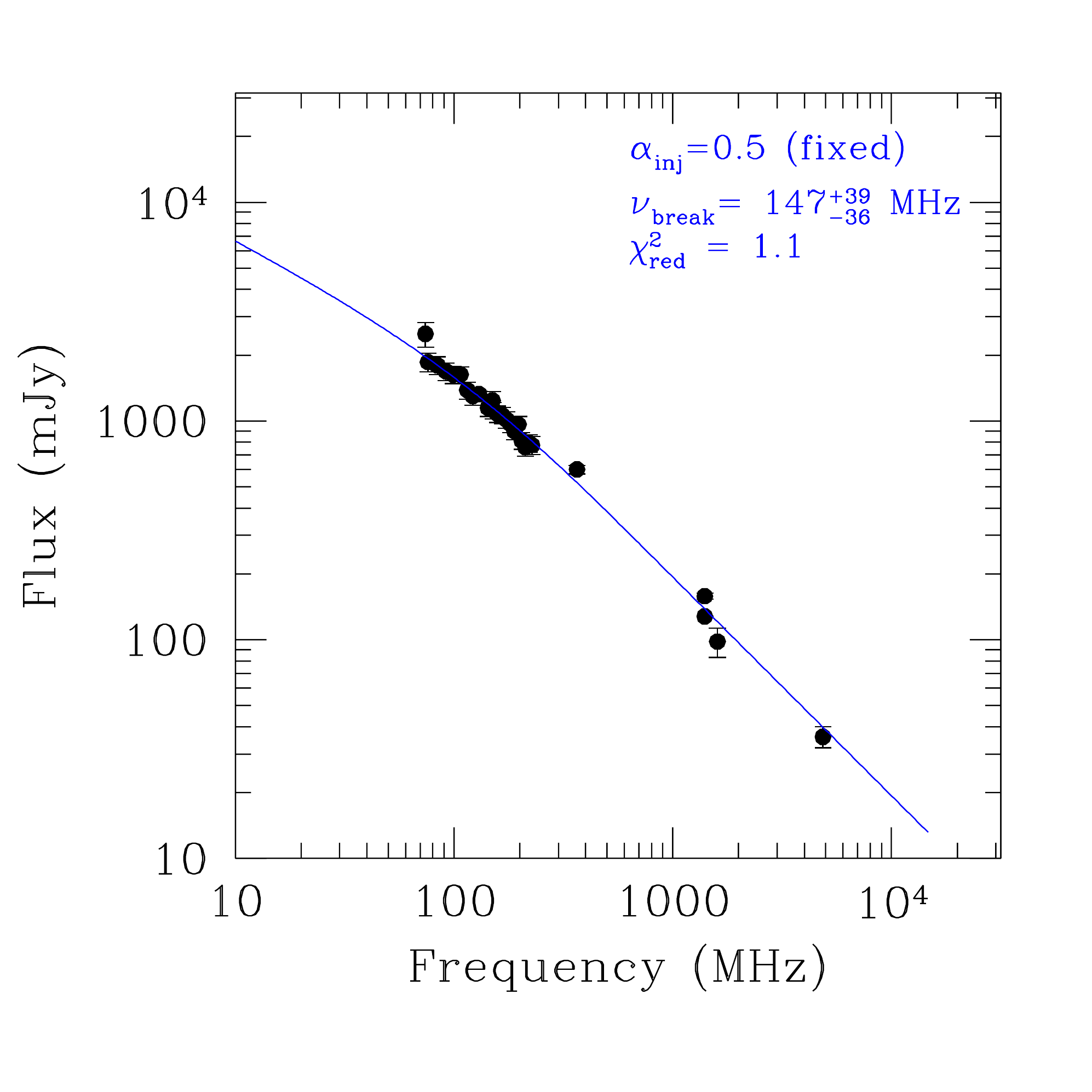}
\caption{\label{fig:radio_spec}Radio spectrum of 1321+045 using data drawn from the VLSSredux, GLEAM, TGSS, TXS, NVSS and FIRST surveys, and our measured VLA and MERLIN flux densities. The line shows the best-fitting continuous injection model described in the text.}
\end{figure}

An alternative approach that allows us to estimate the age of the outer lobes is the radiative age \citep[as described in, e.g.,][]{Parmaetal07,Murgiaetal11,Giacintuccietal12}. The highest energy relativistic electrons in the lobe plasma will radiate away their energy fastest, leading to a break in the radio spectrum, with a steeper spectral index above that break. As the plasma ages, the break will move to lower frequencies, and the frequency of the break can thus be used to estimate the age. \citet{Kunert-Bajraszewskaetal13} found that 1321+045 was well-described by a single power law between 74~MHz and 5~GHz. However, a number of new or improved low frequency survey measurements have become available in the last few years, and we therefore re-examine the integrated radio spectrum. We assembled data from several surveys, including the Very Large Array Low-frequency Sky Survey Redux \citep[VLSSredux,][]{Laneetal14}, Galactic and Extragalactic All-sky Murchison Widefield Array survey \citep[GLEAM,][]{HurleyWalkeretal17}, the TIFR GMRT Sky Survey \citep[TGSS ADR,][]{Intemaetal17}, the Texas Survey of Radio Sources at 365~MHz \citep[TXS,][]{Douglasetal96}, NRAO VLA Sky Survey \citep[NVSS,][]{Condonetal98} and the Faint Images of the Radio Sky at Twenty centimeters survey \citep[FIRST,][]{Beckeretal95} and added our own measured 1.6 and 5~GHz fluxes from the MERLIN and VLA data. Figure~\ref{fig:radio_spec} shows the resulting spectrum. The GLEAM data in particular show a break in the spectrum at low frequencies. We find that the data are well described (reduced $\chi^2$=1.1) by a continuous injection (CI) model with a break at $\nu_{\rm br}$=147$^{+39}_{-36}$~GHz and a spectral index above the break of $\alpha$=1.00$\pm$0.04, in fairly good agreement with the previous measurement and our estimate of the spectral index of the lobes (see Section~\ref{sec:vlba_res}). The injection spectral index of the model was found to be close to $\alpha_{\rm inj}$=0.5, and was fixed at this value during the fit. We note that although there is an apparent discrepancy between the two lowest frequency points in the spectrum (from GLEAM and VLSSredux), this is $<$2$\sigma$ significant taking into account the uncertainties. 

We also tried fitting a CI+off model, in which the jet is assumed to have shut down at t$_{\rm off}$, ceasing to inject particles into the lobes and producing an exponential cutoff in the radio spectrum at high frequencies. \citet{Orientietal10} provide an example of such modelling applied to a GPS source, PKS~1518+047. We found that the CI+off model does not provide a significantly better fit to the spectrum (reduced $\chi^2$ remains 1.1) and that $t_{\rm off}$ is poorly constrained and consistent with zero within 1.1$\sigma$ uncertainties (i.e., with the jets still powering the lobes). The model fits a ratio of the jet shutdown time to the age of the source, $t_{\rm off}$/$t_{\rm s}$ with a best fitting value 0.026$^{+0.013}_{-0.025}$, a break frequency $\nu_{\rm br}$=195$^{+50}_{-83}$~MHz and $\alpha_{\rm inj}$ fixed at 0.5. The frequency of the exponential cutoff, $\nu_{\rm br2}$ is related to the low-frequency break $\nu_{\rm br}$ and timescales via:

\begin{equation}
\nu_{\rm br2} = \nu_{\rm br} \left( \frac{t_{\rm s}}{t_{\rm off}} \right)^2 .
\end{equation}

\noindent For our fitted CI+off model this means the exponential cuttof occurs at a frequency $\nu_{\rm br2}$$\sim$290~GHz, well outside the range of our flux density measurements.

The radiative age is dependent on the break frequency $\nu_{\rm br2}$ and the equipartition magnetic field strength of the source $B_{\rm eq}$, as:

\begin{equation}
t_{\rm rad} = 1590  \frac{B_{\rm eq}^{0.5}}{(B_{\rm eq}^2 + B_{\rm CMB}^2)} [(1+z) \nu_{\rm br}]^{-0.5},
\end{equation}

\noindent where $B_{\rm CMB}$=3.2(1+$z$)$^2$ is the equivalent magnetic field of the cosmic microwave background at $z$, the redshift of the source. Magnetic field strengths are in units of $\mu$G, frequencies in GHz and $t_{\rm rad}$ in Myr.
\citet{Kunert-Bajraszewskaetal13} estimated that the likely magnetic field strength in the radio lobes is high, $\sim$150~$\mu$G. With the break frequency found from the CI model fit, this implies an age of 2.0$^{+0.3}_{-0.2}$~Myr. If we take the CI+off fit instead, treating the jet shut down time as an upper limit, this implies an age for the lobes of $\sim$1.74~Myr, and a jet shut down timescale of $<$0.11~Myr.

\subsection{AGN realignment}
\label{sec:realignment}

Whatever the true age of the lobes, it seems plausible that we are observing two cycles of jet activity. The older cycle was a relatively short outburst, producing the outer lobes. The AGN jets then likely shut down for some period, perhaps as short as tens of thousands of years, perhaps as long as a hundred thousand years. Then in the last few hundred years the jets have restarted, apparently on a quite different axis. These cycles of activity are fairly common in radio sources \citep[sometimes called ``flickering'', e.g.,][]{Maccagnietal20} and in clusters and groups we can in some cases observe evidence of three or even four cycles \citep[e.g.,][]{Randalletal15,Schellenbergeretal21}.

We can consider whether the apparent change in jet axes could be caused by motion of the lobes in the ICM. The lobes will be subject to buoyant forces, which will cause them to rise up the entropy gradient of the ICM. One can imagine scenarios in which, with jets aligned along the major axis of a highly elliptical ICM, buoyant forces might act to move radio lobes along an axis transverse to that of the jets. However, the lobes of 1321+045 are very close to the BCG nucleus, where the entropy distribution appears relatively circular, at least in projection. We see no indication of the strong gradients that would be needed to cause such motion.

Another possibility is that bulk motion of the ICM could move the lobes from their original position. There are many examples of radio lobes affected by motions in the ICM, for example by the sloshing motions associated with minor mergers \citep[e.g., ][]{Kolokythasetal19}. However, the lobes of 1321+045 are in the central few kiloparsecs of the BCG, a location in which they are likely to be protected from any such motion by gas bound to the gravitational potential of the BCG itself. The lobes are also fairly well-centred on the radio core. If they were originally formed along the line of the young jet and moved to their current positions by gas motions, this would involve something like a 90\degree\ rotation of the gas around the AGN which seems implausible. It should also be noted that both buoyant forces and gas motions are typically slow compared to the likely timescale of formation of the radio lobes. In both cases, the lobe motions would be subsonic. The local sound speed in the gas (for kT$\sim$2.8~keV) is $\sim$870\kmps. To reach its current distance from the radio core the east lobe would take $\sim$6.7~Myr moving at the sound speed. Slower gas motions would likely take at least twice as long to move the lobes from the axis of the young jet to their current location.

It therefore seems more likely that the difference in jet axes between the inner jet and outer lobes is caused by a reorientation of the central engine. Several mechanisms can drive such a reorientation: merger or gravitational interaction with a second SMBH, presumably brought into the BCG by a galaxy merger; relativistic precession of the accretion disk and jet caused by a misalignment of the disk and SMBH spin axes; torques applied to the SMBH by a massive, thin accretion disk associated with high accretion rates; or the slower accretion of a significant mass of material whose angular momentum axis does not match that of the SMBH. We see no obvious signs of a recent major merger in the BCG, though they might not be obvious in the available data. We also see no signs of the quasar-mode activity (high optical or infrared luminosity, spectrally hard X-ray emission) expected if a high accretion rate thin disk were present. 

If the jets are aligned relatively close to the line of sight, precession is a possibility. In this case the large change in projected jet axis could be caused by only a small change in true jet axis. An accretion disk misaligned from the SMBH spin axis by a few degrees could be reoriented by relativistic frame dragging \citep[e.g.,][]{McKinneyetal13,Liskaetal18}. The fact that we only see a single-sided jet in the VLBA data could be consistent with such a scenario, since relativistic Doppler boosting would make the approaching jet brighter and the counter-jet fainter (and thus undetected). The difference in brightness and morphology of the two old radio lobes might also be consistent. It is also worth noting that on scales of a few parsecs, acceleration within jets can lead to them showing apparently curved trajectories \citep[e.g.,][]{Jorstadetal17,Listeretal19}. Bearing in mind that our VLBA data do not resolve the structure of the inner jet, if it is aligned close to the line of sight, curvature could lead to a mismatch between its apparent projected axis and the actual path it will follow at larger radii. However, in both cases the angle to the line of sight would need to be small, and from our viewpoint we would expect the accretion disk to be face-on, with little obscuration. This then raises the question of why the AGN nucleus is not detected in the X-ray or via optical emission lines. 

The timescale of the reorientation is too short for slow accretion to significantly change the rotation axis of the SMBH. We do not have an estimate of the SMBH mass, but assuming it to be a few $\times$10$^8$-10$^9$\Msol, the Eddington accretion rate is likely $\lesssim$20\Msolpyr\ \citep{Babuletal13}. For a timescale of 3$\times$10$^4$~yr, even at this high accretion rate, the mass of material accreted would be only $\sim$6$\times$10$^5$\Msol, less than one per cent of the likely black hole mass. For accretion rates well below Eddington, the mass of material which can be accreted in this timescale could not affect the SMBH orientation. Reorientation by this mechanism would also imply a change in the accretion disk axis, with no clear cause for such a change. \citet{Soker18} suggests a model in which gas uplifted and pushed aside by expanding lobes falls back at an angle to the jet axis, causing a change in the disk axis, but the timescales of our source are too short for such infall to have taken place. The cause of the reorientation of the jets is therefore unclear.

\subsection{Cavities and Feedback}
The surface brightness modelling shows no clear evidence of cavities in the ICM. Based on the size of the lobes and the ICM properties in the central bin of the radial spectral profiles, we can estimate the number of counts we would expect the surface brightness to be suppressed by if each lobe is empty of X-ray emitting thermal plasma. Assuming the lobes can be approximated by prolate ellipsoids, we find that for the larger eastern lobe we would only expect about 10 counts (0.5-7~keV) from that volume in our combined image. The lobe region contains 93 counts in our image, spread over 9 pixels. This means that the likely signal from any cavity would only be $\sim$1$\sigma$ significant. We should therefore not be surprised that the cavities are not detected.

Assuming that, in line with other cluster-central radio sources, the radio lobes pushed aside the ICM gas as they inflated, producing cavities filled with relativistic plasma, we can estimate their enthalpy, the combination of the work done on the ICM gas as the cavity is inflated, plus the energy stored in the particles and magnetic field in the lobe. The enthalpy is defined to be 4$pV$ where $p$ is the pressure of the surrounding ICM and $V$ is the lobe volume. We follow the method described by \citet{OSullivanetal11b} in estimating the uncertainty on the lobe volumes, which takes into account the possibility that the apparently prolate lobes may in fact be oblate, and assuming small uncertainties (0.2\arcs) on their radii. Table~\ref{tab:cav} shows the parameters used to estimate lobe enthalpy, including the ICM pressure in the central bin of our radial profile. In total, we estimate the combined enthalpy of the two lobes to be 8.48$^{+6.04}_{-3.56}$$\times$10$^{57}$~erg. This is sufficient to balance radiative cooling in the central 16~kpc of the cluster (the central 2 radial profile bins the lobes overlap) for $\sim$9~Myr, or the entire cooling region for which t$_{\rm cool}<$7.7~Gyr for $\sim$9$\times$10$^5$~yr. 

\begin{table*}
\caption{\label{tab:cav}Parameters used in estimating the enthalpy of the radio lobes. R is the radius of the midpoint of the lobe from the radio core, r$_{\rm maj}$ and r$_{min}$ the semi-major and -minor axes of the ellipses used to approximate the lobes.}
\centering
\begin{tabular}{lcccccc}
\hline
Lobe & R & r$_{\rm maj}$ & r$_{min}$ & Volume & Pressure & Enthalpy \\
 & (arcs) & (arcs) & (arcs) & (10$^{66}$ cm$^3$) & (10$^{-10}$ erg cm$^{-3}$) & (10$^{57}$ erg) \\
\hline
NE & 1.84 & 1.18 & 0.60 & 3.49$^{+3.62}_{-1.75}$ & 3.91$\pm$0.75 & 5.46$^{+5.72}_{-3.03}$ \\
SW & 1.15 & 0.78 & 0.55 & 1.94$^{+1.18}_{-1.12}$ & 3.91$\pm$0.75 & 3.03$^{+1.91}_{-1.88}$ \\
\hline
\end{tabular}
\end{table*}

\citet{Kunert-Bajraszewskaetal13} estimate the magnetic pressure in the radio lobes to be $\sim$9$\times$10$^{-10}$~erg~cm$^{-3}$. This would suggest that the lobes, though perhaps no longer powered by the AGN, are over-pressured by a factor of $\sim$2.3 compared to their environment. This is not implausible given the youth of the source. We therefore caution that our estimates of the enthalpy of the lobes may be under-estimated by a similar factor.

To estimate the mechanical power of the jets we would require the expansion timescale of the radio lobes which, as described in Section~\ref{sec:timescales}, is somewhat uncertain. If we adopt the longer estimated age of $\sim$2.0~Myr, the powers of the two jets would be P$_{\rm jet}$=9.0$\times$10$^{43}$\ergps\ and 5.0$\times$10$^{43}$\ergps\ for the NE and SW lobes respectively. This is at least a factor $\sim$2 less than the bolometric luminosity of the cooling region, 3.13$\pm$0.05$\times$10$^{44}$\ergps. The progress of the tip of the jet would be highly supersonic, and the lateral expansion of the lobes mildly trans-sonic, $\sim$770-840\kmps\ compared to a sound speed in the central 8~kpc of $\sim$660\kmps. We might therefore expect additional energy injection from shocks. If we instead adopt the shortest timescale estimate (3$\times$10$^4$~yr) then we would expect strong shocks to be driven by the lobe inflation. However, since there is no evidence of shocks in the \chandra\ data we cannot make any estimate of the potential heating by this mechanism.

\subsection{ICM cooling}
With its central (isobaric) cooling time of 390$\pm$150~Myr and entropy $<$30\kevcmsq\ within the central 24~kpc (17.1$_{-2.6}^{+2.7}$\kevcmsq\ at 10~kpc), \MaxBCG\ is clearly a strong cool core cluster. Values of the ratio of cooling time to free-fall time (t$_{\rm cool}$/t$_{\rm ff}$) in the range 10-25 have been found to be correlated with the presence of filamentary nebulae and other signs of cooling in cluster cores. We therefore calculate t$_{\rm cool}$/t$_{\rm ff}$ for this system.

\begin{figure}
\centering
\includegraphics[width=8.4cm,bb=30 220 560 725]{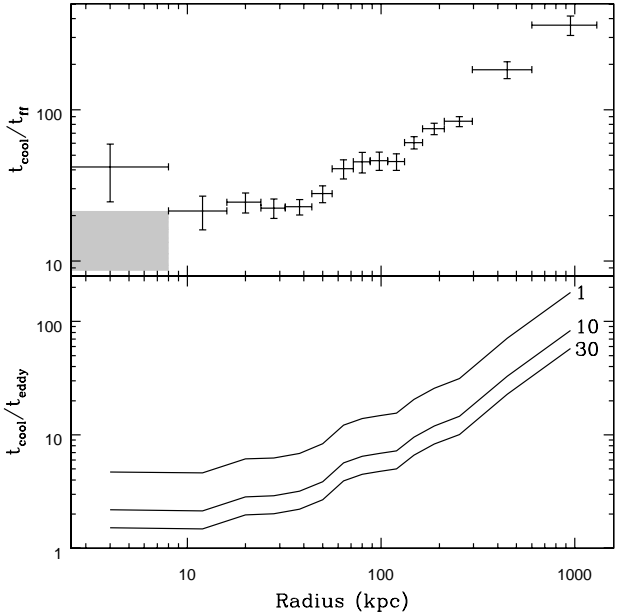}
\caption{\label{fig:tff}Profiles of the ratio of isochoric cooling time to free fall time (estimated from the gas pressure and density profiles, \textit{upper panel}) and eddy time (\textit{lower panel}) in the ICM. The grey shaded region shows an alternate estimate of the free fall time for the inner part of the BCG, based on the stellar velocity dispersion. t$_{\rm cool}$/t$_{\rm eddy}$ profiles are labelled with the injection length scale, $L$ (in kiloparsecs) used in their calculation.}
\end{figure}

Following \citet{Liuetal19}, and assuming a spherical cluster in hydrostatic equilibrium, we can estimate the gravitational acceleration $g$ as

\begin{equation}
    g = \frac{d\Phi}{dr} = \frac{1}{\rho}\frac{dP}{dr}
\end{equation}

\noindent where $\Phi$ is the gravitational potential, $P$ is the ICM pressure, and $\rho$ is the gas density, defined as $\rho$=$1.92$n$_{\rm e}\mu$m$_{\rm p}$ where n$_{\rm e}$ is the electron density, $\mu$ is the mean molecular weight, 0.607, and m$_{\rm p}$ is the proton mass. The free-fall time is then $\sqrt{2r/g}$. Figure~\ref{fig:tff} shows the resulting isochoric t$_{\rm cool}$/t$_{\rm ff}$ profile.

Since the ICM pressure profile is not smooth, we fit it with a generalized NFW model, and calculate $dP/dr$ from the model profile. We also ignore the central bin during the fit, since the pressure value in this bin is probably underestimated. The model therefore likely overestimates the pressure gradient, and therefore the free fall time, in the central bin. \citet{Voitetal15} estimate free-fall times in the centers of giant elliptical galaxies from their stellar velocity dispersions ($\sigma_*$), since t$_{ff}\simeq r/\sigma_*$ in these regions where the stars are the dominant mass component. Using the SDSS stellar velocity dispersion ($\sigma_*$=255$\pm$27\kmps) we estimate the ratio of the (isochoric) cooling time to the free-fall time, t$_{\rm cool}$/t$_{\rm ff}$ to be 15.0$\pm$6.3 in the central 8~kpc bin. This is a little lower than the values estimated from the ICM properties, but still comparable to them. 

We can also consider the ratio of cooling time to the eddy turnover timescale. \citet{Gasparietal18} argue that for their chaotic cold accretion model of AGN fuelling, cold gas condenses from density fluctuations in the ICM, caused by turbulence injected by AGN jets, rising cavities, or other gas motions. Where t$_{\rm cool}$/t$_{\rm eddy}$ approaches unity, condensation can be expected. The eddy turnover time is defined as:

\begin{equation}
{\rm t}_{\rm eddy} = 2\pi\frac{r^{2/3}L^{2/3}}{\sigma_{v,L}}
\end{equation}

\noindent where $L$ is the injection length scale of the turbulence and $\sigma_{v,L}$ is the velocity dispersion of the turbulence at the injection scale. Neither of these two parameters is directly measurable with current data. The only measurement of the (line of sight) ICM velocity dispersion in a cluster to date comes from the \textit{Hitomi} observation of the Perseus cluster \citep{Hitomicollab16}, which found $\sigma_{los}$=164$\pm$10\kmps\ 30-60~kpc from the core. The velocity dispersion of H$\alpha$ and CO nebulae may also be an indicator. In 1321+045, the SDSS measured H$\alpha$ line of sight velocity dispersion is $\sim$183\kmps, with a similar value measured from the [OIII] line. Including a factor of $\sqrt{3}$ to convert from line of sight to three dimensional velocity dispersion, we therefore adopt $\sigma_{v,L}$=317\kmps. The injection length scale $L$ may be determined by the size of the radio lobes (up to 6~kpc) or perhaps by dynamical motions associated with the likely ongoing merger. Since the scale is unclear, we calculated t$_{\rm eddy}$ for $L$=1, 10 and 30~kpc, with the results shown in Figure~\ref{fig:tff}.

The t$_{\rm cool}$/t$_{\rm ff}$ profile, with ratios below 30 out to $\sim$45~kpc, and a central ratio that may be as low as 15, suggests that the ICM in the cluster core may be thermally unstable and thus apt to cool and condense to form atomic or molecular gas. Comparison of the entropy profile, cooling time profile, and minimum t$_{\rm cool}$/t$_{\rm ff}$ value with strongly cooling low redshift clusters in the samples of \citet{Cavagnoloetal09} and \citet{Hoganetal17} shows that \MaxBCG\ has similar thermodynamic properties to these systems. Figure~\ref{fig:deproj} shows comparisons  for entropy and isobaric cooling time; our cluster is at the lower end of the range of profiles for strong cool core clusters, but is not exceptional. The t$_{\rm cool}$/t$_{\rm eddy}$ profiles, which approach unity within 10-50~kpc for injection length scales of 10-30~kpc, support the idea that the ICM may be cooling and condensing to form denser gas phases, provided turbulence is being injected on these scales.  

The SDSS spectrum of the BCG shows both H$\alpha$ and [O\textsc{ii}] emission, and \citet{Liuetal12} classify it as a composite star formation + AGN source. They estimate star formation rates of 3.79 and 5.38\Msolpyr\ based on the H$\alpha$ and [O\textsc{ii}] line strengths, though of course these may be overestimated if the ionized gas is actually located in a filamentary cooling nebula, as in many nearby cooling flow clusters \citep[e.g.,][]{Olivaresetal19}. The H$\alpha$ luminosity of the BCG is 4.5$\times$10$^{41}$\ergps, comparable to the luminosity of nearby clusters such as Abell~1795 \citep{Russelletal17b}, and may be underestimated since it is based on a single fiber spectrum.

The presence of luminous H$\alpha$ emission is certainly consistent with the physical properties of the hot ICM, which are indicative of rapid cooling. If the H$\alpha$ emission is arising from cooled material, we might also expect molecular gas to be present. Our IRAM~30m observations suggest that the BCG is relatively poor in molecular gas for its H$\alpha$ luminosity, but not to an extreme degree. It is quite plausible that we are observing a cluster whose ICM is rapidly cooling, but which has not yet built up the large reservoir of cooled material which is needed to fuel a more sustained AGN outburst. High spatial resolution imaging, or spectral imaging, of the H$\alpha$ emission is probably the most productive next step in determining the state of the system. If an extended filamentary H$\alpha$ nebula is present, it would be a strong indicator of the build up of material cooled from the ICM, and interferometric observation might then be used to determine whether the H$\alpha$:CO ratio is truly unusual. 

\subsection{Evidence of a recent minor merger}
As described in Section~\ref{sec:res}, the X-ray data clearly show that the cluster has an asymmetric surface brightness and temperature distribution. The asymmetric features include the $\sim$6~kpc offset between the BCG nucleus and the peak of ICM X-ray emission, the trail of excess emission extending NNW from the BCG, the large-scale tail of cool, low entropy gas south of the cool core, and the spur of cooler temperatures extending toward the second brightest cluster galaxy. This galaxy, 
SDSS~J132421.40+041918.5, is located 28\arcs/114~kpc to the east of the BCG, and has a redshift $z$=0.26058, 740\kmps\ offset from the BCG. Its SDSS $i$-band Petrosian magnitude is $i$=17.539, compared to $i$=16.676 for the BCG.

Small offsets between BCG and X-ray centroids are quite common in galaxy clusters \citep[see, e.g.,][]{Sandersonetal09,Hudsonetal10,Pasinietal19,Pasinietal21} and can be caused by a number of factors, including dynamical disturbances and the disruption of the core by AGN jets. In this system the radio lobes seem to be too small to have significantly affected the ICM distribution, but the observed structures seem quite consistent with dynamical disturbance of the cluster core by a minor merger.

The southern tail and eastern spur of cool, low entropy gas may be the remnant of the ICM of the infalling subcluster, with SDSS~J132421.40+041918.5 perhaps being its BCG. We can suggest a scenario in which the subcluster fell in from the south, its gaseous halo interacting with the ICM of the main cluster and becoming partly separated from the galaxy population, which moved ahead of the gas, swinging around the cluster core and arcing off to the east. In this scenario the majority of the gas associated with the infalling subcluster is still located south of the main cluster core, while the eastern spur traces the more limited amount of gas still associated with the infalling BCG.

The analysis of the galaxy population by \citet{WenHan13} suggests that \MaxBCG\ is relatively relaxed, and the presence of the strong cool core is a demonstration that any infalling object must have been significantly lower mass than the main cluster. If the $\sim$4~keV temperatures observed in the southern tail are a good approximation of the intruder's ICM temperature, they would suggest a relatively massive system, but mixing and the combination of emission from gas at different temperatures along the line of sight may mean this temperature is biased high. New observations to measure the redshift distribution of the galaxy population of the cluster might be instructive.

\section{Summary and Conclusions}

1321+045 is a rare example of a CSS radio source hosted in the BCG of a galaxy cluster, \MaxBCG. As a young source, it offers an opportunity to study the conditions in the ICM at the time when AGN activity is first triggered. Previous observations have shown the cluster to be X-ray luminous with a pronounced cool core, and confirmed the presence of a radio core with two small lobes. The BCG is a luminous H$\alpha$ source (L$_{\rm H\alpha}$=4.5$\times$10$^{41}$\ergps), though the distribution of the ionized gas is unknown. To investigate this system, we have used a combination of \chandra, VLBA, VLA, MERLIN and IRAM~30m telescope observations, probing the state of the ICM and AGN. We summarize our conclusions below:

\begin{enumerate}
\item The VLBA and MERLIN images suggest that the previously observed 5~kpc-scale radio lobes may no longer be powered by the radio core, which appears to have recently (in the last few hundred years) launched a new $\sim$20~pc jet with a projected axis at right angles to that of the lobes. If the core is still powering the lobes, the jets would need to be oriented close to the line of sight and quite strongly bent, and such an alignment seems unlikely given the lack of detection of the core at X-ray or optical wavelengths. While we cannot place strong constraints on the formation timescale of the radio lobes, spectral aging arguments suggest that, based on a measured break in the radio spectrum at 147$^{+30}_{-36}$~MHz, the lobes are 2.0$^{+0.3}_{-0.2}$~Myr old, while a typical expansion timescale would suggest they may be as young as 3$\times$10$^4$~yr. Although the lobes are too small to be detected in our \chandra\ observation, we estimate their likely enthalpy to be 8.48$^{+6.04}_{-3.56}\times 10^{57}$~erg, sufficient to balance cooling in the surrounding 16~kpc radius volume for $\sim$9~Myr. The reason for the apparent shutdown, reorientation and restart of the AGN jet is unclear, since this appears to be a low luminosity, low accretion rate AGN with no obvious indications of a recent black hole merger.

\item We characterize the ICM of the surrounding galaxy cluster, \MaxBCG, finding that it hosts a strong cool core with low central entropy (8.6$^{+2.17}_{-1.4}$\kevcmsq\ within 8~kpc), short central cooling time (390$^{+170}_{-150}$~Myr), and t$_{cool}$/t$_{ff}$ and t$_{cool}$/t$_{eddy}$ consistent with thermal instability out to $\sim$45~kpc. The azimuthally averaged profiles of ICM properties are similar to those of cool core clusters seen at low redshift. These properties suggest that the ICM is likely to be rapidly cooling, and thus that the H$\alpha$ emission likely arises from material cooled from the ICM.

\item Our IRAM~30m observations do not detect emission from molecular gas in either the CO(1-0) or CO(3-2) lines, and we place 3$\sigma$ upper limits on the molecular gas mass of M$_{\rm mol}\leq$7.7$\times$10$^9$\Msol\ and $\leq$5.6$\times$10$^9$\Msol\ from the two lines respectively. These limits are a factor 1.3-2.8 below the molecular gas mass expected from the known correlation between H$\alpha$ and CO emission, but given the scatter in that relationship, this is not necessarily a significant deviation. Taking the non-detection at face value, it may suggest that the ICM is still in the process of building up a reservoir of cooled gas which will in future power further AGN outbursts.

\item We find that \MaxBCG\ has probably undergone a minor merger in the recent past, which has left a trail of cool, low entropy gas extending from the south, past the cool core and BCG, and pointing to the second brightest galaxy in the cluster, SDSS~J132421.40+041918.5. While a detailed analysis of the interaction would require a redshift survey of the cluster, it seems likely that this galaxy was the BCG of an infalling subcluster whose gas has been stripped by interaction with the ICM to form the tail we observe in the X-ray spectral maps.
\end{enumerate}

To further explore the state of the radio lobes and young jet, and the factors that may have triggered AGN outbursts in this source, would require additional observations. High spatial resolution radio observations could determine the spectrum of the older radio lobes, giving a better constraint on their age and physical properties. On milli-arcsecond scales, the detection of a counter-jet might give some idea of the orientation of the current AGN axis with respect to the line of sight, while measurement of the spectrum of the core and jet might allow us to confirm whether they can be classed as a CSS or GPS source separate from the radio lobes. High resolution H$\alpha$ imaging or imaging spectroscopy could answer the question of how the H$\alpha$ emission is distributed, and therefore of its origin; the presence of an extended filamentary nebula of ionized gas would be a strong indicator of cooling from the ICM. Optical spectroscopy of the nucleus might also help determine whether the AGN is a genuinely low-luminosity, low accretion rate system, or whether heavy obscuration is hiding a more luminous core. However, our \chandra\ observations may have answered our most basic question about this system. Despite the youth of the radio source, the properties of the cluster ICM are similar to those of clusters in the local universe, suggesting that at least in this case, the AGN has not been triggered by runaway cooling.

\medskip
\noindent\textbf{Acknowledgments}\\
The authors thank the anonymous referee for their constructive and helpful comments, and the staff of the \chandra\ X-ray Observatory, VLBA and IRAM~30m telescope for their help with the observations used in this work.
The authors gratefully acknowledge the support for this
work provided by the National Aeronautics and Space Administration (NASA)
through \chandra\ Award Number GO0-21112X issued by the \chandra\ X-ray
Center, which is operated by the Smithsonian Astrophysical Observatory for
and on behalf of the National Aeronautics Space Administration under contract NAS8-03060. 
MKB acknowledges support from the 'National Science Centre, Poland' under grant no. 2017/26/E/ST9/00216.
Basic research in radio astronomy at the Naval Research Laboratory is supported by 6.1 Base funding.
The National Radio Astronomy Observatory is a facility of the National Science Foundation operated under cooperative agreement by Associated Universities, Inc. This work is based on observations carried out under project number 208-19 with the IRAM 30m telescope. IRAM is supported by INSU/CNRS (France), MPG (Germany) and IGN (Spain). This research has made use of the NASA/IPAC Extragalactic Database (NED), which is funded by the National Aeronautics and Space Administration and operated by the California Institute of Technology.

\facilities{CXO, VLBA, VLA, IRAM:30m, MERLIN}

\software{CIAO \citep[v4.12][]{Fruscioneetal06}, Sherpa \citep[v4.12][]{Freemanetal01}, XSpec \citep[12.11.0h][]{Arnaud96}, AIPS \citep{VanMoorseletal96}, GILDAS \citep{Pety05,GILDAS13}}

\bibliographystyle{aasjournal}
\bibliography{../paper}

\end{document}